\documentclass{aa} 

\usepackage{graphicx}
\usepackage{txfonts}
\usepackage{pdflscape}
\usepackage{comment}
\usepackage{orcidlink}
\usepackage[switch]{lineno}
\usepackage{hyperref}
\newcommand{\Halpha}{\ion{H}{$\alpha$}}
\newcommand{\Hbeta}{\ion{H}{$\beta$}}
\newcommand{\FeIII}{\ion{Fe}{III}}
\newcommand{\SII}{\ion{S}{II}}
\newcommand{\NII}{\ion{N}{II}}
\newcommand{\CaII}{\ion{Ca}{II}}
\newcommand{\OI}{\ion{O}{I}}
\newcommand{\MgI}{\ion{Mg}{I}}
\newcommand{\NaID}{\ion{Na}{ID}}

\begin{document}

 \title{Searching for late-time interaction signatures in Type Ia supernovae from the Zwicky Transient Facility}

 \author{Jacco H. Terwel \orcidlink{0000-0001-9834-3439} \inst{1,2}
 \and Kate Maguire \orcidlink{0000-0002-9770-3508} \inst{1} 
 \and Georgios Dimitriadis \orcidlink{0000-0001-9494-179X} \inst{1}
 \and Mat Smith \orcidlink{0000-0002-3321-1432} \inst{3,4}
 \and Simeon Reusch \orcidlink{0000-0002-7788-628X} \inst{5,6}
 \and Leander Lacroix \orcidlink{0000-0003-0629-5746} \inst{7,8}
 \and Lluís Galbany \orcidlink{0000-0002-1296-6887} \inst{9,10}
 \and Umut Burgaz \orcidlink{0000-0003-0126-3999} \inst{1}
 \and Luke Harvey \orcidlink{0000-0003-3393-9383} \inst{1}
 \and Steve Schulze \orcidlink{0000-0001-6797-1889} \inst{8}
 \and Mickael Rigault \orcidlink{0000-0002-8121-2560} \inst{4}
 \and Steven L. Groom \orcidlink{0000-0001-5668-3507} \inst{11}
 \and David Hale \inst{12}
 \and Mansi M. Kasliwal \orcidlink{0000-0002-5619-4938} \inst{13}
 \and Young-Lo Kim \orcidlink{0000-0002-1031-0796} \inst{3}
 \and Josiah Purdum \orcidlink{0000-0003-1227-3738} \inst{12}
 \and Ben Rusholme \orcidlink{0000-0001-7648-4142} \inst{11}
 \and Jesper Sollerman \orcidlink{0000-0003-1546-6615} \inst{14}
 \and Joseph P.~Anderson \orcidlink{0000-0003-0227-3451} \inst{15,16}
 \and Ting-Wan Chen \orcidlink{0000-0002-1066-6098} \inst{17}
 \and Christopher Frohmaier \orcidlink{0000-0001-9553-4723} \inst{18}
 \and Mariusz Gromadzki \orcidlink{0000-0002-1650-1518} \inst{19}
 \and Tom\'as E.~M\"uller-Bravo \orcidlink{0000-0003-3939-7167} \inst{9,10}
 \and Matt Nicholl \orcidlink{0000-0002-2555-3192} \inst{20}
 \and Shubham Srivastav \orcidlink{0000-0003-4524-6883} \inst{20}
 \and Maxime Deckers \orcidlink{0000-0001-8857-9843} \inst{1}
 }

 \institute{School of Physics, Trinity College Dublin, The University of Dublin, Dublin 2, Ireland\\
 \email{terwelj@tcd.ie}
 \and Isaac Newton Group (ING), Apt. de correos 321, E-38700, Santa Cruz de La Palma, Canary Islands, Spain
 \and Department of Physics, Lancaster University, Lancs LA1 4YB, UK
 \and Université de Lyon, Université Claude-Bernard Lyon 1, CNRS/IN2P3, IP2I Lyon, 69622 Villeurbanne, France
 \and Deutsches Elektronen Synchrotron DESY, Platanenallee 6, D-15738 Zeuthen, Germany
 \and Institut für Physik, Humboldt-Universität zu Berlin, D-12489 Berlin, Germany
 \and LPNHE, CNRS/IN2P3, Sorbonne Université, Université Paris-Cité, Laboratoire de Physique Nucléaire et de Hautes Énergies, 75005 Paris, France
 \and Oskar Klein Centre, Department of Physics, Stockholm University, Albanova University Center, SE-106 91, Stockholm, Sweden
 \and Institute of Space Sciences (ICE-CSIC), Campus UAB, Carrer de Can Magrans, s/n, E-08193 Barcelona, Spain.
 \and Institut d’Estudis Espacials de Catalunya (IEEC), E-08034 Barcelona, Spain.
 \and IPAC, California Institute of Technology, 1200 E. California Blvd, Pasadena, CA 91125, USA
 \and Caltech Optical Observatories, California Institute of Technology, Pasadena, CA 91125
 \and Division of Physics, Mathematics, and Astronomy, California Institute of Technology, Pasadena, CA 91125, USA
 \and Oskar Klein Centre, Department of Astronomy, Stockholm University, Albanova University Center, SE-106 91, Stockholm, Sweden
 \and European Southern Observatory, Alonso de C\'ordova 3107, Casilla 19, Santiago, Chile
 \and Millennium Institute of Astrophysics MAS, Nuncio Monsenor Sotero Sanz 100, Off. 104, Providencia, Santiago, Chile
 \and Graduate Institute of Astronomy, National Central University, 300 Jhongda Road, 32001 Jhongli, Taiwan
 \and School of Physics and Astronomy, University of Southampton, Highfield, Southampton SO17 1BJ, UK
 \and Astronomical Observatory, University of Warsaw, Al. Ujazdowskie 4, 00-478 Warszawa, Poland
 \and Astrophysics Research Centre, School of Mathematics and Physics, Queens University Belfast, Belfast BT7 1NN, UK
 }

 \date{Received XXX; accepted YYY}

 
 \abstract
 {The nature of the progenitor systems and explosion mechanisms that give rise to Type Ia supernovae (SNe Ia) are still debated. The interaction signature of circumstellar material (CSM) being swept up by the expanding ejecta can constrain the type of system from which it was ejected. However, most previous studies have focused on finding CSM ejected shortly before the SN Ia explosion, which still resides close to the explosion site resulting in short delay times until the interaction starts. We use a sample of 3\,627 SNe Ia from the Zwicky Transient Facility that were discovered between 2018 and 2020 and search for interaction signatures greater than 100 days after peak brightness. By binning the late-time light curve data to push the detection limit as deep as possible, we identify potential late-time rebrightening in three SNe Ia (SN 2018grt, SN 2019dlf, and SN 2020tfc). The late-time optical detections occur between 550 and 1450\,d after peak brightness, have mean absolute \textit{r}-band magnitudes of $-$16.4 to $-$16.8 mag and last up to a few hundred days, which is significantly brighter than the late-time CSM interaction discovered in the prototype, SN 2015cp. The late-time detections in the three objects all occur within 0.8 kpc of the host nucleus and are not easily explained by nuclear activity, another transient at a similar sky position, or data quality issues. This is suggestive of environment or specific progenitor characteristics playing a role in the production of potential CSM signatures in these SNe Ia. Through simulating the ZTF survey, we estimate that $<$0.5 per cent of normal SNe Ia display late-time ($>100$ d post peak) strong \Halpha-dominated CSM interaction. This is equivalent to an absolute rate of $8_{-4}^{+20}$ to $54_{-26}^{+91}$ Gpc$^{-3}$ yr$^{-1}$ assuming a constant SN Ia rate of $2.4\times10^{-5}$ Mpc$^{-3}$ yr$^{-1}$ for $z \leq 0.1$. Weaker interaction signatures of \Halpha\ emission, more similar to the strength seen in SN 2015cp, could be more common but are difficult to constrain with our survey depth.
 }

 \keywords{supernovae: general: -- supernovae: individual: SN 2018grt, SN 2019ldf, SN 2020tfc-- circumstellar matter}

 \maketitle
%

\section{Introduction}
Type Ia supernovae (SNe Ia) span a range of peak absolute magnitudes that can be standardised using properties of their light curves around peak \citep[e.g.][]{Phillips_rel, Phillips_rel2}. However, besides normal SNe Ia, there are also events that are spectroscopically and/or photometrically different, creating their own subclasses and that can not be standardised in the normal way. One such subclass are those that are thought to be interacting with circumstellar material (CSM), called SN Ia-CSM events. The first SN Ia-CSM reported was SN 2002ic that had \Halpha~and \Hbeta~emission in its spectrum around peak, which was suggested to originate from interaction with CSM \citep[][]{02ic_H_det, Hamuy_02ic}. The optical light curve of SN 2002ic was found to be declining much more slowly than was expected for normal SNe Ia \citep{02ic_slow_decay}, and found to be consistent with about 1.3 $M_\odot$ of CSM being present around the progenitor system \citep{single_degen_CSM_gen}. 

It has been suggested that there is a link between the Ia-CSM and 91T subclasses \citep[bright SNe Ia with a distinct pre-peak spectroscopic evolution, showing a blue pseudo-continuum with two strong \FeIII~absorption multiplets instead of intermediate mass element lines, and predominantly occurring in younger stellar populations;][]{Filippenko_91t,91T} due to similarities in their peak brightness and spectroscopy before the start of the interaction \citep{Ia-CSM_and_91T_connection}. In some cases, a 91T-like SN could start interacting with CSM hundreds of years after the explosion, like as has been suggested for Kepler's SN \citep{Kepler_91T, Kepler_CSM}.

Currently the Ia-CSM subclass contains several dozen members \citep{2005gj, Ia-CSM_Silverman, Ia-CSM_BTS}. In all cases, the CSM interaction started within about two months of explosion, significantly altering the light curve and leading to the (re-)classification as a Ia-CSM after spectroscopic confirmation. The amount and duration of interaction varies quite significantly from one object to another. While some mainly feature an unusually slow decline rate (e.g. SN 2018gkx and SN 2020xtq; \citealt{Ia-CSM_BTS}), other objects have a plateau for several 100\,d before starting to fade away slowly (e.g. SN 2020aekp; \citealt{Ia-CSM_BTS}). 

SN 2011km (PTF11kx) is a well observed Ia-CSM event, and shows signs of a complex CSM consisting of multiple shells with which the SN ejecta interact. \citet{ptf11kx} show that its photometry is similar to 91T-like objects before the CSM interaction begins. They explain the geometry of the system using a symbiotic nova progenitor system.
Another interesting Ia-CSM is SN 2020eyj, which \citet{Kool_He_CSM} present as the first detection of a Ia-CSM interacting with He-rich material. Up until then all members of the subclass had shown strong \Halpha~emission and weak He signatures. SN 2020eyj, however, showed little to no H present in the CSM. This was also the first time a SN Ia was detected in the radio. Non-detections in normal SNe Ia suggest a clean environment for the ejecta to expand in, while in this case there is a lot of material present. Considering this, \citet{Kool_He_CSM} suggest the progenitor system to have been a He star and white dwarf (WD) binary.

The progenitor systems and mechanisms responsible for triggering the explosions of SNe Ia are still not clear. The progenitor system could be single degenerate, with a main sequence or evolved non-degenerate star accompanying a WD \citep{Whelan_classical_Ia_mod, Nomoto_single_degenerate}. Mass is transferred from the companion to the WD until the explosion is triggered. In the double degenerate scenario, both stars are WDs and merge or interact \citep{Iben_Double_degenerate, Webbink_Double_degenerate}. The different progenitor scenarios can be separated into two categories of explosion models. In classical models the WD comes close to or reaches the Chandrasekhar mass (M$_\text{Ch}\sim 1.4~\text{M}_\odot$, \citealt{Chandrasekhar_lim}) by accreting matter from the secondary star before exploding \citep{Whelan_classical_Ia_mod}, possibly through a delayed detonation \citep{Kholov_Del_det, Mazzali_common_mechanism}. The second category assumes a situation where the explosion is triggered in a lighter WD, resulting in a sub-Chandrasekhar mass explosion. This can be done in e.g. the double detonation models where an accreted layer of surface material ignites and explodes, compressing the WD which ignites the core causing a second explosion \citep{Taam_ddet, Livne_ddet, Shen_ddet, Fink_ddet}, or in mergers with the core of an evolved star in the core-degenerate scenario \citep{Kashi_core_deg}. Other models involve the collision or violent merger of two WD \citep{Rosswog_merger, Pakmor_merger, Pakmor_merger2}.

The CSM identified to be present in Ia-CSM could be created by different mechanisms depending on the progenitor system, and its composition depends on the type of donor star present in the system. In the single degenerate scenario, CSM rich in hydrogen can be created by a WD generating a fast wind, blowing away a part of the material it received from the mass transfer \citep{single_degen_CSM_gen}. In the double-degenerate scenario, part of the tidally disrupted secondary WD becomes unbound from the system, creating (H-poor) CSM and may be able to produce detectable signatures depending on the time between the tidal disruption and the SN Ia explosion \citep{Double_degen_CSM_gen}. When the SN ejecta reach and start to sweep up the CSM surrounding the system, the interaction is revealed as an additional source of light which brightens and alters the light curve of the SN. Depending on the distance of the CSM to the explosion site, there may be a significant delay between the explosion and the start of the interaction.

In the effort to systematically search for late-time CSM interaction, \citet{2015cp} looked at old ($\geq 1$ year) SNe using the \textit{Hubble Space Telescope (HST)}. They focused their search on subclasses, such as 91T, that are associated with CSM interaction \citep{Ia-CSM_and_91T_connection}. Out of 72 targets, only ASASSN-15og and SN 2015cp were found to show late-time CSM interaction. ASASSN-15og is a Type IIn SN with detected CSM interaction around peak, and was used as a control object. SN 2015cp had been classified as a 91T-like SN Ia, without signs of CSM interaction around peak. This showed that CSM interaction may start much later after the explosion, and may be systematically missed due to SNe Ia not being actively followed at these phases. From a progenitor point of view this means that material can be ejected from the system prior to the explosion, potentially giving it time to travel further before being caught up by the SN ejecta.

\cite{GALEX_Late_CSM} used archival UV-band data from the \textit{Galactic Evolution Explorer (GALEX)} to look for late-time CSM interaction in SNe Ia. Out of a sample of 1080 SNe Ia, 4 were detected in the UV near peak, but none showed signs of late-time CSM interaction. They show that this type of CSM interaction is rare, occurring between 500 to 1000\,d after the initial discovery of the SN in $<5$ per cent of the SNe Ia at a strength similar to SN 2015cp, and a decreasing percentage as the interaction gets stronger.

With today's large sky surveys such as the Zwicky Transient Facility (ZTF, \citealt{ZTF_Surveys_Scheduler, ZTF_Science_Objectives, ZTF_Instrumentation, ZTF_Observing_System}) and Asteroid Terrestrial-impact Last Alert System (ATLAS, \citealt{ATLAS}), transient events are discovered and followed automatically until they fade below the detection limit. This strategy is extremely efficient in finding and cataloguing transients and is a reliable method to find rare subclasses and interesting objects, assuming their defining features can be identified before they fade away. However, depending on the phase of a Ia-CSM when interaction begins (potentially greater than one year post peak), it may be systematically missed as the SN may no longer be actively followed, or the detections could be close to the detection threshold and thus not necessarily be associated with the original SN.

ZTF has covered the entire northern sky above declination of $-30^\circ$ every 2 -- 3 nights in three broad optical bands (\textit{gri}) up to limiting magnitudes of $\sim$20.5 mag since early 2018. A survey this deep and extensive both in space and time coverage may have detected late-time CSM interaction in SNe Ia that has gone unnoticed due to being close to the detection limit. We attempt to push this limit as much as possible by binning the post-SN observations together for each confirmed SN Ia observed with ZTF before December 2020. While binning data will reduce our temporal resolution, it is traded for deeper detections and upper limits.

In Section \ref{data}, we introduce the sample we use in our search for optical signals for late-time CSM interaction in the ZTF data stream. In Section \ref{analysis}, we present our custom pipeline for identification of late-time flux excesses and set up a simulation to test it and estimate the detection efficiency of our pipeline. Section \ref{results} shows the result of running our sample through our pipeline, and provides further investigation on some interesting objects. These results are discussed in Section \ref{discussion}, and we conclude in Section \ref{conclusion}. A flat $\Lambda$CDM cosmology for H$_0 = 67.7$\,km\,s$^{-1}$\,Mpc$^{-1}$ and $\Omega_m = 0.310$ \citep{Planck18VI} is assumed where required.

\section{Data}
\label{data}
Our aim is to look for late-time ($>$100\,d after peak brightness) signatures of CSM interaction in the largest sample of SNe Ia to date. This has been obtained by the ZTF. We are particularly interested in events that appear to be normal SNe Ia from their spectra and light curves around peak but may display signs of late-time interaction, as seen in SN 2015cp \citep{2015cp}. Our starting sample is 3\,627 events that were discovered by ZTF from March 2018 to October 2020 (hereafter the ZTF data release 2, ZTF DR2). Each event was spectroscopically classified as a SN Ia or one of its sub-classes. An overview of the ZTF DR2 will be presented in Rigault et al.~(in prep.), including the sample definition, properties, and use for cosmology. In this study, since we are searching for likely rare signatures of interaction in the ZTF light curves, we are as inclusive as possible in our sample definition and include all SNe Ia in the DR2 covering a redshift out to $z = 0.289$. 

\subsection{ZTF light curve data}
\label{lc_data}
ZTF observes in three optical bands $gri$ on a 2 -- 3 day cadence. Reference images, mainly made using observations at the start of ZTF, are subtracted from the science images using the \textsc{zogy} image subtraction algorithm \citep{ZOGY} to produce difference images. We use forced photometry at the transient location on the difference images using \textsc{ztffps} \citep{ztffps} to get a measure of the observed flux at each epoch. This includes non-detections before each was first detected and after each SN has faded below detection limits. 

 Light curve quality cuts on specific light curve points are applied as in Rigault et al (in prep.). We do not correct the light curves for Milky Way extinction in our initial analysis but do consider it when focusing on specific objects of interest in Section \ref{results}. 

Another approach for extracting photometry at the transient location is by using Scene Modeling Photometry \citep[SMP;][]{Holtzmann_SMP}. We extract SMP for a few selected objects of interest in Section \ref{results} to test if the identified late-time detections are independent of our approach. When using SMP, one has to define an `off' time and and `on' time. The observations taken during the `off' time are used to create a model of the region, or scene, where the SN occurs. This is then used as a template during the `on' time to calculate and remove host contributions to the photometry, leaving just the transient itself (Lacroix et al.~in prep.). The advantage of this method is the significantly lower uncertainty in the model compared to the difference imaging technique, allowing us to find fainter detections. Since we assume that a signal from late-time interaction could occur at any point after the SN, we define the `off' time as everything up to shortly before the SN explodes, and the `on' time as everything after this moment.

\subsection{Baseline correction}
 Issues in the construction of the reference images such inaccurate flat-fielding or artefacts in the images that are subsequently co-added can result in a systematic offset in the forced photometry light curve made using different images. The technique of baseline correction is used to correct for this \citep{Yao_baseline_corr,Miller_baseline_corr}. 

To estimate the necessary baseline correction, we calculated the weighted mean of the flux of all data points up to 40 days before the estimated SN peak (which is assumed to be the highest flux detection deemed real), and subtracted it from the light curve. Baseline corrections are done separately for each combination of band ($gri$), field (telescope pointing), and rcid (part of the camera, which is arranged in 4$\times$4 charge-coupled devices (CCDs) with four readout channels each, giving a total of 64 readout channels) as each of these combinations uses different, unique reference images. To be able to apply a baseline correction, at least two observations are needed. If this is not possible all observations with that band, field, rcid combination are removed.

Since we are interested in post-SN detections, our baseline correction method using only pre-SN data differs from the one used in Rigault et al.~(in prep.) with both pre- and post-SN data. A comparison between the methods found that for most objects our corrections agree with the ones used in Rigault et al. (in prep.) within the uncertainties. This is as expected as objects with late-time flux excesses are expected to be rare, meaning that the two baseline correction methods should give the same result for most objects. 

If there is insufficient data to perform a baseline correction, the relevant data (based on field, filter and rcid) is removed from the light curve. If this includes data around the peak position, the peak position in the light curve may change and therefore, the position of the peak is recalculated.

\begin{table}
 \centering
 \caption{The initial sample size and its reduction in each step of the analysis process.}
 \begin{tabular}{lrr}
  \hline
  Criterion & Removed & Objects left\\
  \hline
  Initial DR2 sample & - & 3\,627\\
  No photometry at 100+ days & 109 & 3\,518\\
   No `robust' late-time detections$^a$ & 2\,952 & 566\\
  Presence of SN Ia tail$^b$ & 432 & 134\\
  Removed on visual inspection$^c$ & 101 & 33\\
  No late-time CSM interaction$^d$ & 30 & 3\\
  \hline
 \end{tabular}
 \label{obj_breakdown}
\begin{flushleft}
$^a$For a `robust detection', at least four positive detections (two or more adjacent bins with $\ge$5$\sigma$) are required out of the 16 possible combinations of bin size (25, 50, 75, 100 d), and the starting position of the bin varied by 25, 50 or 75 per cent of the bin size. \\
$^b$We tested for the presence of a radioactive tail of the SN Ia as described in Sec.~\ref{tail_removal} and removed those where this was the most plausible explanation. \\
$^c$Each remaining light curve was inspected using \textsc{snap} (Sec.~\ref{snap}) for possible issues causing late-time detections. See Sec.\,~\ref{results} for a discussion of the reasons events were removed.\\
$^d$For each remaining light curve we checked in detail if the late-time detections could be explained without CSM interaction starting at late times.
\end{flushleft}
\end{table}

\section{Analysis}\label{analysis}

To systematically search for objects with late-time flux excesses, a custom pipeline was developed. In Section \ref{pipeline}, we describe how the late-time photometry for each object is binned to reach deeper magnitude limits, as well as the scheme used to select objects with robust, significant detections. In Section \ref{tail_removal}, we identify and remove bright nearby SNe Ia whose late-time detections are due to the SN radioactive decay tail. In Section \ref{snap}, we describe our method of visually inspecting images of potentially interesting sources, and in Section \ref{simulation}, we detail our use of \textsc{simsurvey} to simulate SNe Ia with late-time interaction signatures. Table~\ref{obj_breakdown} shows our sample size after each step of analysis that is discussed in the subsequent sections.

\subsection{Binning \& filtering program}
\label{pipeline}

After pre-processing, the late-time observations are binned in phase to push the detection limit beyond the limit of the individual observations. We define late-time observations as being at least 100 days after the estimated date of SN peak brightness in the observer frame. We remove all SNe Ia that have no data in any band beyond this phase. The exact choice of 100 days is arbitrary but was chosen as balance between minimising spurious detections due to the light curves being dominated by SN light at earlier times and maximising the phase range over which interaction can be searched for. 

Binning of the light curves is performed on each band separately. Larger bins are better for pushing the magnitude limit as deep as possible, they sacrifices temporal sensitivity. To balance the time sensitivity and magnitude limit, we use bins with widths of 100, 75, 50, and 25 days. To make sure that the placement of the bin edges does not affect our results, we repeat the binning four times for each bin size, while adjusting the starting epoch of the binning by decreasing the size of the first bin by 25 per cent in each iteration. This results in a total of 16 trials for each band. 

\begin{figure}
 \centering
 \includegraphics[width=\columnwidth]{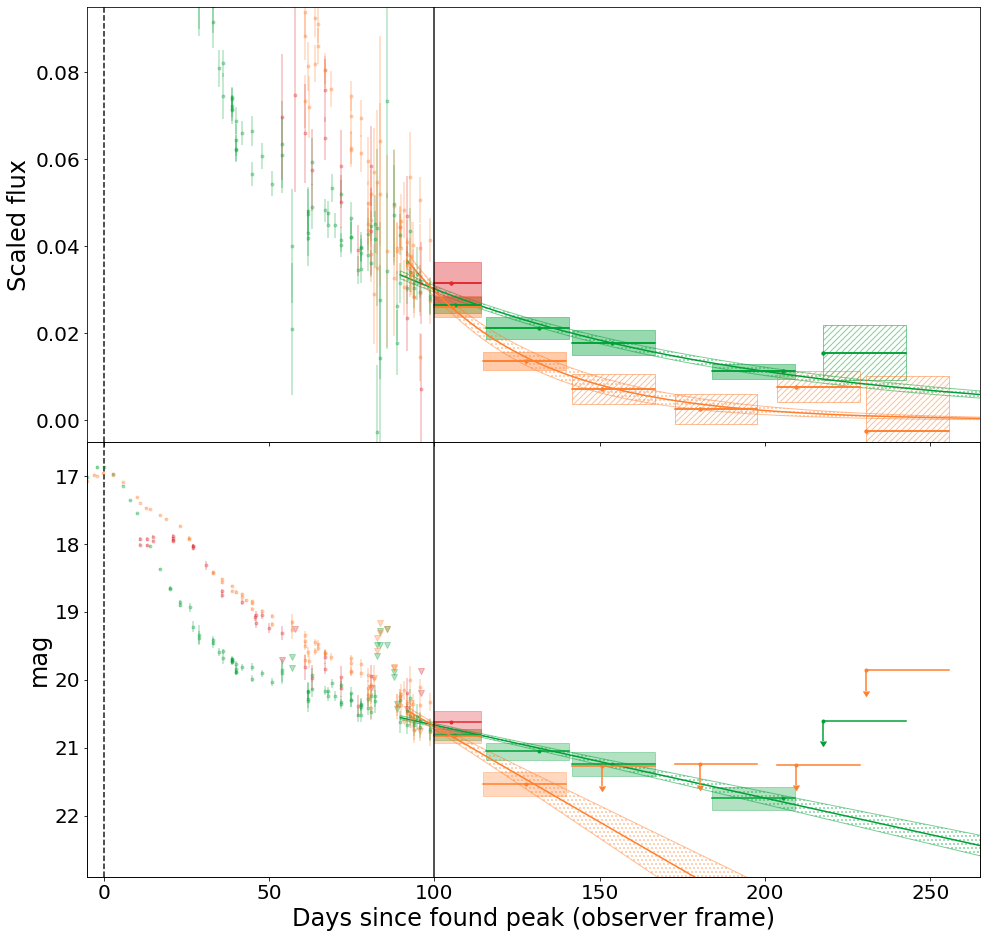}
 \caption{The first 250 days of the \textit{gri}-band light curves of SN 2019hbb in flux scaled to the peak flux (top panel) and magnitude (bottom panel) space. 100 days after the estimated peak date (vertical black dashed line) binning starts (marked with a vertical black solid line), using 25\,d bins. The $g$-, $r$- and $i$-bands are shown green, orange, and red, respectively. Before 100 days, we show the unbinned detections with their uncertainties (coloured circles) and non-detections (inverted triangles). After 100 days we show the bins as horizontal lines to show their size, a circle to show their mean value, and the shaded region showing the 1$\sigma$ uncertainty (dashed regions are non-detections). A bin is deemed a non-detection if the flux $f<5\sigma_f$. The $5\sigma$ magnitude limit is calculated and shown as a downward arrow. In both the $g$- and $r$-bands, the first bin is a detection and there are multiple adjacent bins with detections, triggering the tail-fitting procedure (see Section \ref{tail_removal}). The resulting tail fits are shown in the green and red lines, respectively, with their 1$\sigma$ uncertainties as hashed regions. The half-life times are $t_{1/2,g} = 70 \pm 6$\,d ($\chi^2_\text{red} = 0.6$) and $t_{1/2,r} = 27 \pm 4$\,d ($\chi^2_\text{red} = 1.4$). This tail is therefore deemed to be a normal SN Ia tail.}
 \label{bin_showcase}
\end{figure}

Each bin starts at the phase position of a data point to avoid empty bins. A gap in the data that is larger than the bins being used can cause the bins after the gap to always be placed in the same location despite the size modifications of the first bin. To avoid this we re-apply this modification of adjusting the size of the first bin after the data gap to trial different start positions. Lastly, if a bin would only contain one or two points, and the phases of these points occur no later than 10 per cent of bin size of the previous bin (e.g. if the bin size is 25 d, they would occur within 2.5\,d of the end of the bin), the previous bin is increased in size to include these points. An example of the bin placement is shown in Fig.~\ref{bin_showcase}.

For each bin the weighted mean and uncertainty of the observed flux are calculated. When binning flux measurements taken from difference images, the uncertainty of the reference images used in the difference imaging procedure has to be considered, as this will limit the depth of the binned observations \citep{ref_uncert} and this is added in quadrature to the weighted uncertainty of the binned flux. 

After the binning procedure each light curve has undergone 16 trials per band across four bin sizes and four bin placements. An attempt in a specific trial is considered significant if it has two or more adjacent bins with $\ge$5$\sigma$ detections. A late-time detection is considered `robust' if at least four out of 16 attempts have significant detections suggesting that the detections are insensitive to bin placement and/or size. We make this choice of `robust' detection in at least four attempts to ensure that we are not dominated by spurious detections but that we can still detect a long but faint signal that can only be picked up in the four trials involving the largest (100 day) bins. 

\subsection{Removal of SN Ia radioactive tail detections}
\label{tail_removal} 
For some nearby SNe Ia, the normal light curve tail powered by the radioactive decay of $^{56}$Co $\rightarrow$ $^{56}$Fe is still visible at the phases investigated here ($>$100 d after the peak), possibly triggering false positives in our pipeline. To test if the fading tail is the reason for detections after 100 d, we checked if the detections follow a declining power law in flux space consistent with that of a normal SN Ia tail. We take all bins, normalise to the brightest data point, and fit a declining power law. To ensure that the tail matches the earlier data points, we include the unbinned observations between 60 and 100 d after the peak, making sure that if there are N bins only the latest N/2 unbinned detections are used to ensure that the fit focuses on the bins and not the unbinned points.

A successful fit of a declining normal SN Ia tail has a reduced chi-square of the fit of $\chi^2_\text{red, fit} < 5$ and fitted half-life of $t_{1/2}$ with uncertainty $\sigma_{t1/2}$, satisfying $t_{1/2} - 5\sigma_{t1/2} \leq 50$ d. We chose a threshold of 50\,d as \citealt{Georgios_11fe} showed that this is the approximate decay time scale for a normal SN at these phases. A fit with a high $\chi^2_\text{red, fit}$ value could have failed due to bad or uncertain data, or due to the late-time detections not following a power law decay. Fits with a $t_{1/2}$ significantly larger than that of a normal SN Ia tail suggest an additional luminosity source contributing to the light curve at these phases. Figure~\ref{bin_showcase} shows an example where this tail fitting procedure determines the late-time detections in an object to be a normal declining SN Ia tail. 432 SNe Ia that were flagged as having late-time detection are discounted from further discussion because their light curves can be explained by a normal fading SN Ia tail.

\subsection{SuperNova Animation Program (SNAP)}
\label{snap}
 After performing the binning and filtering, and removing SNe Ia with contamination from the SN radioactive tail, we are left with 134 SNe Ia with robust late-time detections (see Table \ref{obj_breakdown}). Since the binning and filtering programs are designed to handle a large quantity of light curves and cannot be tailored specifically to suit the peculiarities of a single object, it is possible that there are objects remaining with issues in the data or data processing (e.g.~cosmic rays, bad subtractions), resulting in false positive detections. Therefore, we have manually checked the difference imaging to search for potential issues. To do this efficiently, we made \textsc{snap}\footnote{\url{https://github.com/JTerwel/SuperNova_Animation_Program}}.

Using \textsc{ztfquery} \citep{ZTFquery}, \textsc{snap} takes all difference images of the requested position in the sky during the requested time period(s) in the requested band(s) and shows them in chronological order in an animation. At the start of each animation the reference images in all bands are shown. The program can show the image in grey-scale, a three-dimensional wire-frame representation of the intensities measured per pixel, the averaged values along both axes of the image, the observation date and duration, the peak and mean pixel values of the shown region, the last spectrum taken before the currently shown image, and highlight the resulting forced photometry point in the light curve corresponding to the plotted images.

Using \textsc{snap}, issues in the difference images were identified, including SN ghosts (the SN is visible in the reference image, leaving a negative imprint in difference images after it has faded), cosmic rays, and bad pixels (NaN, or a large negative number). Variability of a separate source can also be seen, which when close by can contaminate the forced photometry at the SN location, e.g.~an Active Galactic Nucleus (AGN).

\subsection{ Simulated interaction recovery fractions }
\label{simulation}
To make sure the binning program works as expected and estimate its detection efficiency in finding late-time signals, an observing campaign was simulated using \textsc{simsurvey} \citep{simsurvey, simsurvey_main}, a python package designed to simulate large scale time domain surveys, such as ZTF. To successfully simulate an observing campaign, the program needs to be told what, when, where, and how something is observed, and under what conditions. For this, we need a model of the SN Ia-CSM that is going to be observed, an explosion rate as a function of redshift, and a time range for these explosions to occur. We also require an observing log specifying which part of the sky is being observed at a specific time, the length of the observations, and the weather conditions during the observations. Lastly, we require details of the camera that is used to carry out the observations. The SN model needs to be in a similar format to the \textsc{sncosmo} models (a \textsc{python} package made for supernova cosmology, \citealt{sncosmo}). This means we need a set of spectra over the entire phase range, all having the same wavelength spacing and range. Since no such model exists, we built our own model as described in the next section.

\subsubsection{The interacting SN model}
\label{model_description}

SN 2011fe was chosen as the template of a normal SN Ia as it is well observed and close by (in M101 at a distance of 6.4 Mpc, \citealt{M101_cep_dist}). Optical spectra of SN 2011fe were obtained between phases of $-$18 to 1017\,d relative to the peak. We made a custom model using \textsc{sncosmo} and spectra found on WISeREP \citep{wiserep}\footnote{\url{https://wiserep.weizmann.ac.il}}, which are listed in Table \ref{11fe_sources}. The spectra were flux calibrated to match the observed coeval broadband magnitudes. Spectra up to 45\,d after the peak were flux calibrated using a \textsc{salt2} \citep{salt2} fit of the \textit{PTF48g} and \textit{PTF48R}-band photometry \citep{PTF_1, PTF_2}, which is used to estimate the flux in the $g$- and $r$- bands at these phases. Spectra between 45 and 400\,d are flux calibrated using a cubic spline interpolation of photometry in the \textit{PTF48g}- and \textit{PTF48R}-bands. The interpolation extends up to 600\,d after the peak in the \textit{PTF48R}-band, there is no \textit{PTF48g} photometry used between 400 and 600\,d after the peak. Three interpolated photometry points from the \textsc{salt2} fits were used as anchor points to connect these two parts of the calibration. \citet{Georgios_11fe} show that there is a slight kink in the light curve tail around 600\,d after the peak. This is replicated in the model by calibrating all spectra more than 600\,d after the peak using a cubic spline interpolation of photometry from the Large Binocular Telescope \citep[LBT;][]{LBT} in the Bessel \textit{R}-band \citep{Shappee_11fe}.

After flux calibration, the spectra were dereddened to remove dust extinction effects, using the \citet{ccm89_extinction_law} extinction law with $A_V = 0.04$ mag \citep{extinction_Av}. The spectra were rebinned, and any wavelength region that was not covered in all spectra was removed as required by \textsc{simsurvey}. Lastly, the model is corrected for distance, redshift and time dilation. The resulting model is that of a normal SN Ia, which exploded at a distance of 10 pc without any dust between the source and observer.

 The best late-time detection of CSM in a normal/91T-like SN Ia was for SN 2015cp, where \Halpha\ emission was identified in its spectra at 664 d after light-curve peak \citep{2015cp}. To model potential CSM interaction signals similar to that of SN 2015cp, we added a narrow \Halpha~line with a Gaussian profile to the SN 2011fe model. Due to the rareness of interaction in otherwise normal SNe Ia, we do not have good constraints on the diversity of interaction signatures and we have simulated a broad parameter space. The interaction was chosen to start at 100, 200, 300, or 500 days after the peak, last for 100, 300, or 500 days, and have a similar brightness to the observed signal in SN 2015cp \citep{2015cp}, as well as 10 times weaker or 10 times stronger than it. All possible combinations of these values are used, and a simulation without any interaction is also used as a control test, giving a total of 37 simulations. 

\begin{figure}
 \centering
 \includegraphics[width=\columnwidth]{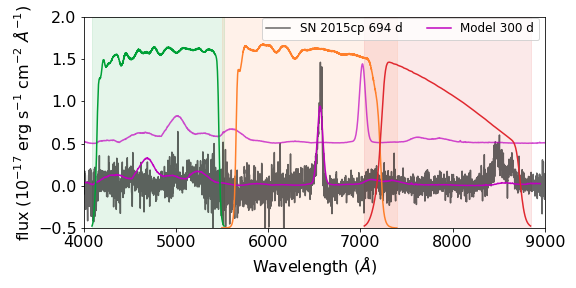}
 \caption{The model spectrum at 300 days (SN 2011fe with the added \Halpha~line) is shown in magenta overlaid on a rest-frame spectrum of SN 2015cp at 694 days in grey. The model flux has been scaled to the distance of SN 2015cp for comparison. The green, orange, and red shaded regions are the bandwidths of the $g$-, $r$-, and $i$-bands, respectively. The transmission profiles are plotted in the same colours for each band. The model is also shown shifted to $z = 0.07$ (and offset up in flux), where the \Halpha~line has just started to be in the $i$-band.}
 \label{mod_15cp_comp}
\end{figure}

\begin{figure}
 \centering
 \includegraphics[width=\columnwidth]{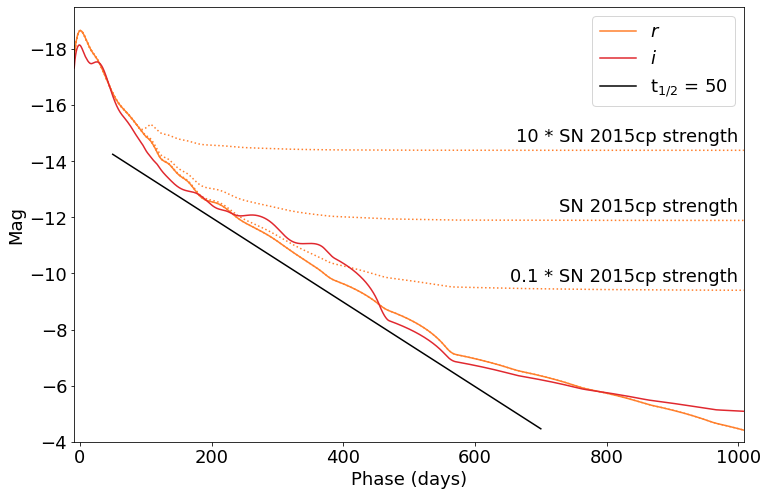}
 \caption{The $r$ (orange) and $i$ (red) absolute magnitude light curves of the SN 2011fe model used in the simulations in the ZTF bands as a function of phase from rest-frame \textit{B}-band peak \citep{spec_HST}. The bumpiness in the models are because the underlying \textsc{sncosmo} model class interpolates in flux space but fails to find an exponential decay. The added rest-frame CSM interaction model based on \Halpha~emission (starting at a phase of 100 d) is shown with dotted lines for the \textit{r}-band. Once the interaction becomes the dominant source, it smooths out the bumps from the underlying tail. The black line shows a radioactive decay with t$_{1/2}=50$ d, typical of a declining normal SN Ia tail. }
 \label{11fe_mods}
\end{figure}

Figure~\ref{mod_15cp_comp} shows an example of our model spectra at 300\,d and SN 2015cp at 694\,d in the rest frame \citep{2015cp}. It also shows the model redshifted to $z = 0.07$, where the \Halpha~line is partly shifted into the $i$-band. Figure~\ref{11fe_mods} shows the absolute magnitude \textit{ri}-band light curves of the SN 2011fe model in the rest frame, as well as light curves with different strengths of \Halpha~emission. Our interaction model will only generate a late-time interaction signal in the $r$-band (or $i$-band at $z > 0.06$) as we only added \Halpha\ emission line. This is enough to test the binning program but is likely too simple to reflect the actual late-time signal seen in SN 2015cp (which also showed \OI\ and \CaII\ in the restframe $i$-band) or potential other events.

\subsubsection{Simulating the observing campaign}
\label{sim_obs}
By specifying the object to observe, as well as the telescope details and observation schedule, an observation campaign can be performed using \textsc{simsurvey} resulting in a collection of observed light curves. For a deeper explanation of \textsc{simsurvey} we refer the reader to \cite{simsurvey_main}. The parameters used as input are listed in Appendix \ref{sec:simsurvey_inputs} . In each \textsc{simsurvey} run, $10^5$ SNe Ia are simulated to produce observed light curves and meta-data such as redshift, observed peak date, etc. To ensure that the SNe Ia are similar at peak to those recovered, we require that the SN Ia light curves must have at least three detections of $\geq$5$\sigma$ and are brighter that 19 magnitude at peak. This reduces the sample $\sim$ 40,000 objects per simulation. These are sent through the late-time excess detection pipeline (Section \ref{pipeline}) as if they were real observed light curves to determine the recovery efficiency. 

The volumetric rate used as input in the simulations favours more distant SNe, which results in very few SNe at extremely low redshift values and hence larger uncertainties. To mitigate this, we split $0\leq z\leq 0.015$ into bins of size 0.001 and simulated an additional 100 SNe in each bin using the same parameters as in the original simulations. Introducing these additional events does not impact the recovery efficiencies because we are comparing the number of recovered events relative to the input number in each redshift bin and therefore, are insensitive to the input rate of events.

\subsubsection{Simulated interaction recovery}
\label{simulated_reco}
\begin{figure}
 \centering
 \includegraphics[width=\columnwidth]{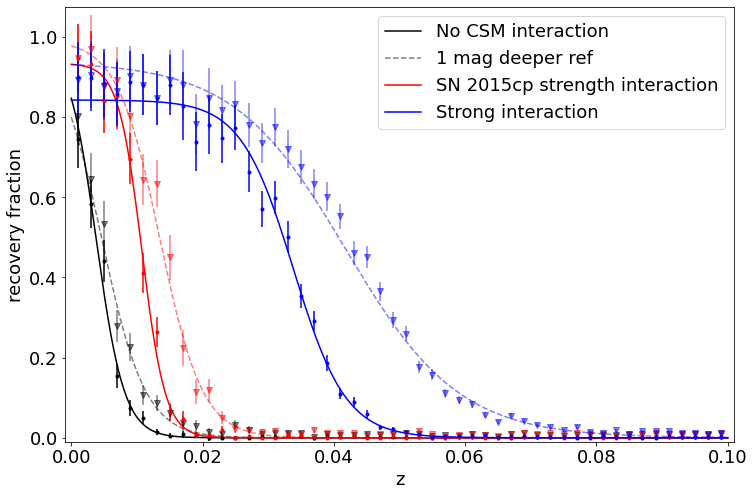}
 \caption{Fraction of SNe Ia for one of our simulations (interaction occurring between 500 -- 1000\,d after the peak) where the CSM interaction was recovered per redshift bin of size $0.002$. The simulations are shown for interaction strengths of zero (grey), similar to SN 2015cp (red), and 10 times stronger than SN 2015cp (blue). In the simulation without CSM interaction, the recovery fraction should be interpreted as the fraction of false positives. The simulations with normal ZTF quality reference images are shown with dots and fitted sigmoid functions with solid lines. Simulations where one magnitude deeper reference images were assumed are shown in triangles, with their fitted sigmoid functions in dashed lines.}
 \label{recov_fracs}
\end{figure}

Our aim is to determine from our simulations how many SNe Ia with signatures of late-time interaction similar to that of SN 2015cp would have been detected by our pipeline. For each of the simulations, we binned the SNe based on their redshift and looked at the fraction of SNe that were reported by the pipeline to show late-time excesses. Figure~\ref{recov_fracs} shows the recovery fractions as a function of redshift for an example simulation when the interaction starts at 500\,d and lasts for 500 d for simulations of no CSM interaction, late-time interaction with the same strength as SN 2015cp, and interaction 10 times as strong as SN 2015cp (`strong interaction'). As expected the recovery fraction drops off with increasing redshift for both the SN 2015cp equivalent strength and the interaction that is 10 times stronger, with the strong interaction recoverable out to a higher redshift. 

The recovery fraction of the simulations with CSM interaction does not reach 100 per cent in the lowest redshift bins. This is because the radioactive tail of these SNe Ia tend to be bright out to hundreds of days after explosion. Therefore, depending on the cadence and uncertainties of the simulated photometry, the SN light can dominate over the CSM interaction and the CSM interaction signal does not alter the shape of the SN decay tail enough to be flagged as CSM interaction. 

In the simulations without CSM interaction, the recovery fraction is non-zero at small redshifts, meaning that some objects are falsely identified as having late-time excess. For these very bright and high signal-to-noise SN Ia light curves, our decaying tail model for normal SNe Ia proves to be too simple. Our analysis pipeline detects real deviations of the SN light curve evolution from our simple decay tail model. This only occurs at the lowest redshifts and nearby SNe Ia are rare, with only 0.6\% of our observed SN Ia sample at $z\leq0.01$. This means that contamination of our sample due to normal SNe Ia tails that cannot be fit by our simplified tail fit model is very low. 

We have fitted a sigmoid function to the recovery fractions of each simulation, using the total amount of objects in each bin as its weight (Fig.~\ref{recov_fracs}). A sigmoid function is an oversimplification (the recovery fraction is underestimated at the low redshifts) but it allows us to easily estimate the redshift limit where CSM interaction can be recovered. We define our redshift limit where CSM interaction can be recovered as $z_{50}$, the redshift where 50 per cent of the SN interactions are recovered. These values are listed for all simulations in Table \ref{sim_z50_results}.

 As discussed in Section \ref{model_description}, we have simulated 36 models with interaction signatures starting at 100, 200, 300 and 500 d post peak, lasting for 100, 300 and 500 d, and with strengths the same as SN 2015cp, 10 times weaker and 10 times stronger. We also simulated a model without any late-time CSM interaction. For the models with an interaction strength similar to SN 2015cp \citep{2015cp}, when the interaction is short and early (starting 100 days after the peak and lasting for 100 days), the interaction can not be distinguished from a normal SN Ia decaying light curve and the recovery fraction is as low as the no interaction simulation. When the interaction is longer, or if it starts later, the light curve flattens enough to be identified as deviating from a normally declining SN Ia tail. This pushes the redshift boundary where 50 per cent of the interaction would be recovered by ZTF to $z_{50} = 0.0105 \pm 0.0003$ for the longest and latest interaction (500 - 1000 days after peak).

If the CSM interaction is 10 times weaker than that of SN 2015cp, the decaying SN Ia tail generally dominates over the interaction signature and the light curve shows little deviation from a normal decaying SN Ia tail. Even in the best case scenario of the longest and latest CSM interaction simulation, the 50 per cent recovery thresholds lies at $z_{50} = 0.0050 \pm 0.0005$. In the simulations where the interaction is 10 times stronger compared than that of SN 2015cp, the shortest and earliest interaction (lasting from 100 to 200 days after peak) has $z_{50} = 0.0091 \pm 0.0014$. For the longest and latest interaction, the the 50 per cent recovery rate is at $z_{50} = 0.0323 \pm 0.0004$.

\subsubsection{Impact of reference image depth}
\label{impact_refdepth}
The mean limiting magnitude of the ZTF reference images is $\sim 21.8$ mag and as discussed in \citep{ref_uncert}, this is the limiting factor for recovering faint signals from binned light curve data. To test the improvement of deeper reference images, the assumed limiting magnitude was changed to be 0.5 and 1 mag deeper. The recovery fraction for one magnitude deeper is shown for comparison in Fig.~\ref{recov_fracs}. As expected, deeper reference images allows the interaction signatures to be detected to higher redshift, although the increases in $z_{50}$ values are modest (see Table \ref{sim_z50_results}). For example, for the latest onset and longest interaction duration interaction, $z_{50}$ increased from $0.0323 \pm 0.0004$ to $0.0407 \pm 0.0009$.

\section{Results}
\label{results}

We have run our custom detection pipeline on the ZTF DR2 light curves in the same way as it was performed on the simulated light curves in Section \ref{simulation}. In 1932 light curves, nothing is detected in any of the 16 trials discussed in Section \ref{pipeline}, in 432 light curves the late-time detections are attributed to declining SN Ia tails, and in 1020 light curves the detections were not considered robust (<4 successful trials). These are the three largest cuts in our sample, as can be seen in Table\,\ref{obj_breakdown}, and leave us with 134 light curves that pass the pipeline. In Section\,\ref{results_summary}, we describe the light curves of these events and discuss how some light curves fit into known classes of events (e.g. known Ia-CSM, late-time SN Ia tail detections). In Section \ref{Additional_tests}, we describe the additional tests that were performed on the remaining promising 10 events to determine if their late-time excesses are due to CSM interaction or other scenarios.

\subsection{Initial summary of detected events}
\label{results_summary}

As can be seen in Table \ref{obj_breakdown}, the result of the pipeline is a list of 134 objects that require visual inspection after passing the detection cuts of positive $>$5$\sigma$ detections in adjacent light curve bins in at least four of the 16 bin size and placement combinations. In 47 of these cases, by visual inspection we identify that an incorrect baseline caused false positives. In five cases, the peak date estimation failed and estimated the peak to be over 100 days before the actual SN explosion. Because of this the SN itself was detected as a late-time signal. Furthermore, in 29 cases there was evidence of the host galaxy being improperly subtracted or showing signs of activity, which interfered with the forced photometry at the SN location. Finally, in 20 cases the tail fit test was unable to show the nature of the tails due to various reasons (e.g. the fits did not converge properly or there was a gap in the observations while the tail was visible). After this step, 33 objects were remaining in the sample.

 \begin{landscape}
  \begin{table}
   \centering
   \caption{The list of objects that passed the initial visual inspections.}
   \begin{tabular}{llccccccccccccc}
    \hline
    Name & IAU name & Redshift & Type$^a$ & Peak & Peak & Excess & Excess & Excess & Group$^b$ \\
    &&&& MJD & mag.& phase (d) & band & mag. \\
    \hline
    ZTF18aaykjei  & SN 2018crl & 0.0968 $\pm$ 0.0004 & Ia-CSM & 58297.3 & 18.40 $\pm$ 0.04  & 100 -- 400 & $r$ & 20.1 -- 21.9 & Known Ia-CSM \\
    ZTF18abuatfp  & SN 2018gkx & 0.1367 $\pm$ 0.0004 & Ia-CSM & 58382.1 & 18.71 $\pm$ 0.05  & 100 -- 475 & $gri$ & 19.6 -- 21.5 & Known Ia-CSM \\
    ZTF18actuhrs  & SN 2018evt & 0.02442 $\pm$ 0.00001 & Ia-CSM & 58476.5 & 16.19 $\pm$ 0.01 & 100 -- 525 & $gri$ & 16.6 -- 21.8 & Known Ia-CSM \\
    ZTF19aaeoqst  & SN 2019agi & 0.0594 $\pm$ 0.0004 & Ia-CSM & 58511.5 & 18.46 $\pm$ 0.04  & 100 -- 450 & $gri$ & 19.2 -- 21.9 & Known Ia-CSM \\
    ZTF19abidbqp  & SN 2019ibk & 0.0402 $\pm$ 0.0002 & Ia-CSM & 58688.5 & 18.56 $\pm$ 0.05  & 100 -- 1125 & $gr$ & 19.1 -- 21.5 & Known Ia-CSM \\
    ZTF19acbjddp  & SN 2019rvb & 0.1832 $\pm$ 0.0004 & Ia-CSM & 58790.1 & 18.90 $\pm$ 0.06  & 100 -- 400 & $gr$ & 20.4 -- 22.0 & Known Ia-CSM \\
    ZTF20aatxryt  & SN 2020eyj & 0.0294 $\pm$ 0.0004 & Ia-CSM & 58939.2 & 17.24 $\pm$ 0.01 & 100 -- 450 & $gr$ & 19.0 -- 21.8 & Known Ia-CSM \\
    ZTF20abbbsfs  & SN 2020kre & 0.13530 $\pm$ 0.00001 & Ia-CSM & 58998.2 & 19.09 $\pm$ 0.04  & 175 -- 425 & $gr$ & 19.8 -- 21.4 & Known Ia-CSM \\
    ZTF20abmlxrx  & SN 2020onv & 0.0940 $\pm$ 0.0004 & Ia-CSM & 59052.4 & 17.85 $\pm$ 0.01  & 100 -- 500 & $gr$ & 18.8 -- 21.6 & Known Ia-CSM \\
    ZTF20abqkbfx  & SN 2020qxz & 0.0968 $\pm$ 0.0004 & Ia-CSM & 59094.3 & 18.18 $\pm$ 0.04  & 100 -- 450 & $gri$ & 19.5 -- 21.9 & Known Ia-CSM \\
    ZTF20accmutv  & SN 2020uem & 0.043 $\pm$ 0.001  & Ia-CSM & 59173.5 & 16.38 $\pm$ 0.01 & 100 -- 525 & $gr$ & 17.4 -- 21.4 & Known Ia-CSM \\
    ZTF20aciwcuz  & SN 2020xtg & 0.06118 $\pm$ 0.00001 & Ia-CSM & 59189.5 & 17.45 $\pm$ 0.02  & 100 -- 500 & $gri$ & 18.0 -- 21.9 & Known Ia-CSM \\
    ZTF20acyroke  & SN 2020aeuh & 0.122 $\pm$ 0.004  & Ia-CSM & 59217.4 & 19.01 $\pm$ 0.06 & 100 -- 250 & $r$ & 19.8 -- 20.9 & Known Ia-CSM \\
    \hline
    ZTF18aasdted  & SN 2018big & 0.01815 $\pm$ 0.00001 & Ia-norm & 58268.4 & 15.64 $\pm$ 0.01 & 450 -- 550 & $gr$ & 20.4 -- 21.5 & Sibling  \\
    ZTF19aaysiwt  & SN 2019hnt & 0.0926 $\pm$ 0.0004 & Ia  & 58651.2 & 18.46 $\pm$ 0.05  & 525 -- 625 & $gr$ & 20.5 -- 21.7 & Sibling  \\
    ZTF19acihfxz  & SN 2019tjz & 0.055 $\pm$ 0.003  & Ia-norm & 58795.1 & 18.00 $\pm$ 0.03  & 950 - 1050 & $r$ & 19.4 -- 20.7 & Sibling  \\
    ZTF20abzetdf  & SN 2020tft & 0.071 $\pm$ 0.002  & Ia-norm & 59113.5 & 18.2 $\pm$ 0.1  & 725 -- 800 & $r$ & 19.4 -- 20.2 & Sibling  \\
    ZTF20acehyxd  & SN 2020uvd & 0.0346 $\pm$ 0.0005 & Ia  & 59129.3 & 18.72 $\pm$ 0.03  & 300 -- 350 & $r$ & 20.2 -- 21.5 & Sibling  \\
    \hline
    ZTF19aatlmbo  & SN 2019ein & 0.0072 $\pm$ 0.0001  & Ia-norm & 58617.2 & blinded$^c$    & 100 -- 425 & $r$ & 18.9 -- 21.8 & Kinked tail \\
    ZTF20abqvsik  & SN 2020rcq & 0.00262 $\pm$ 0.00001 & Ia-norm & 59144.5 & blinded$^c$    & 100 -- 400 & $i$ & 16.5 -- 22.2 & Kinked tail \\
    ZTF20abrjmgi  & SN 2020qxp & 0.00364 $\pm$ 0.00001 & 18byg-like & 59088.1 & blinded$^c$    & 100 -- 375 & $r$ & 17.9 -- 21.8 & Kinked tail \\
    ZTF20abwrcmq  & SN 2020sck & 0.01644 $\pm$ 0.00001 & Iax & 59099.4 & 16.25 $\pm$ 0.01 & 100 -- 450 & $gri$ & 19.1 -- 21.9 & Kinked tail \\
    ZTF20achlced  & SN 2020uxz & 0.00850 $\pm$ 0.00001 & Ia-norm & 59142.4 & blinded$^c$    & 100 -- 400 & $gr$ & 17.1 -- 21.9 & Kinked tail \\
    \hline
    ZTF19acwrqtv  & SN 2019vzf & 0.059 $\pm$ 0.004  & Ia  & 58829.1 & 18.00 $\pm$ 0.07  & 150 -- 400 & $gr$ & 20.7 -- 21.8 & Other -- AGN$^d$   \\ 
    ZTF20aahptds  & SN 2020awr & 0.07656 $\pm$ 0.00001 & Ia-norm & 58888.5 & 18.54 $\pm$ 0.07 & 300 -- 1050 & $i$ & 20.2 -- 20.5 & Other -- data issue$^d$  \\
    ZTF20aazwuin  & SN 2020kzd & 0.082 $\pm$ 0.001  & Ia-norm & 58997.4 & 18.86 $\pm$ 0.03  & 250 -- 850 & $gr$ & 21.3 -- 22.0 & Other -- data issue$^d$  \\
   \hline 
    ZTF18abtqevs$^*$ & SN 2018grt & 0.042 $\pm$ 0.003  & Ia-norm & 58372.3 & 18.56 $\pm$ 0.02 & 1350 -- 1450 & $r$ & 21.1 -- 21.3 & Other   \\
    ZTF19aanyuyh  & SN 2020pkj & 0.02453 $\pm$ 0.00001 & Ia-norm & 59060.4 & 16.90 $\pm$ 0.01& 100 -- 175 & $r$ & 20.8 -- 21.1 & Other   \\
    ZTF19abfvhlx$^*$ & SN 2019ldf & 0.057 $\pm$ 0.003  & Ia-norm & 58686.5 & 17.90 $\pm$ 0.04 & 1050 -- 1225 & $ri$ & 20.1 -- 21.1 & Other   \\
    ZTF19ablekwo  & SN 2019mse & 0.088 $\pm$ 0.004  & Ia-norm & 58715.4 & 18.37 $\pm$ 0.02 & 450 -- 700 & $gri$ & 19.8 -- 20.6 & Other   \\
    ZTF19abzwtiu  & SN 2019rqn & 0.075 $\pm$ 0.003  & Ia-norm & 58760.3 & 18.62 $\pm$ 0.03 & 950 -- 1050 & $r$ & 20.6 -- 21.4 & Other   \\
    ZTF20aaifyfx  & SN 2020alm & 0.05997 $\pm$ 0.00001 & Ia  & 58873.5 & 18.10 $\pm$ 0.02 & 750 -- 1025 & $gri$ & 19.9 -- 21.9 & Other   \\
    ZTF20abjfufv$^*$ & SN 2020tfc & 0.031 $\pm$ 0.001  & Ia-norm & 59116.3 & 17.22 $\pm$ 0.01& 550 -- 800 & $gri$ & 18.9 -- 21.4 & Other   \\
    \hline
  \end{tabular}
  \label{33_list}
  \begin{flushleft}
  *Our three final objects with suggested detections of late-time interaction. \\
  $^a$Type is based on spectral classifications (Rigault et al.,~in prep.), `Ia-CSM' are those interacting with CSM that is observed around peak, `Ia-norm' are those that are most consistent with a normal SN Ia, `Ia' are those without a sub-classification but are consistent with being a SN Ia, SN 2020qxp is 18byg-like \citep{de_2019_18byg}, and SN 2020sck is a Iax \citep{2020sck_Iax}.\\
  $^b$The objects are split into four groups: Ia-CSM that were previously known, identified siblings, SNe Ia that are detected due to their early-time tail fits showing a kink (`Kinked tail'), and `Other' category that includes 10 events with potential flux excesses, including three (SN 2019vzf, SN 2020awr and SN 2020kzd) that are subsequently ruled out (see Section \ref{Additional_tests}). \\
  $^c$The peak magnitudes of four SNe Ia in the sample are blinded because of their planned use in H$_0$ constraints (Rigualt et al.,~in prep.).\\
  $^d$SN 2019vzf is ruled out as a true excess due to AGN variability at the position, SN 2020awr the excess was only detected in the \textit{i} and was ruled out by the scene modelling analysis, and SN 2020kzd was detected in the \textit{gr} bands on a complex galaxy environment and not detected in the scene modelling analysis (see Section \ref{sec:scene_modelling}). 
  \end{flushleft}
 \end{table}
\end{landscape}

Further details of these remaining 33 SNe Ia are shown in Table \ref{33_list} and can be split up into four main groups: i) known Ia-CSM events, ii) transient `siblings', where a second transient event occurs near the identified SN Ia causing its light to (partially) be picked up during forced photometry at the first SN location, iii) nearby objects whose tail could not be fitted by our simple model, and iv) objects that do not fall in the first three groups. In the following sections, we describe the first three of these groups that are not due to potential CSM interaction at late times.

\begin{figure*}
 \centering
 \includegraphics[width=17cm]{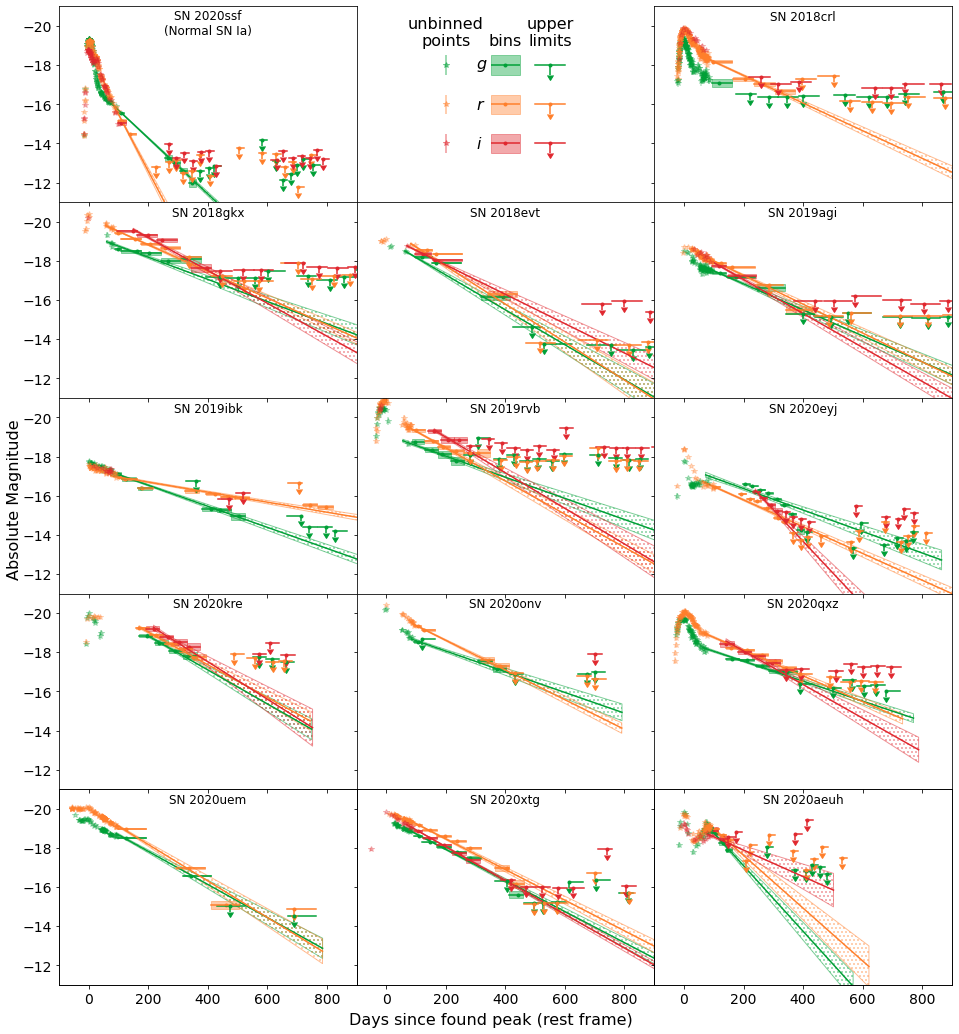}
 \caption{Binned late-time observations of the recovered known SNe Ia-CSM. All objects are shown in absolute magnitude and over the same time range for easy comparison. All objects are detected beyond 100 days after the peak without using the binning technique. We do not show these individual data points to increase readability. The tail fits are shown as solid lines with the hashed region denoting their 1$\sigma$ uncertainties. For comparison, SN 2020ssf (ZTF20abyptpc) in the top left corner is a normal SN Ia with a normally declining tail with $t_{1/2,g} = 53 \pm 1$ days and $t_{1/2,r} = 26 \pm 1$ days. The fitted tails for the SNe Ia-CSM are significantly shallower.}
 \label{known_CSM_plots}
\end{figure*}

\subsubsection{Known Ia-CSM}
\label{known-iacsm}
The first group are the 13 known Ia-CSM, defined as those objects that already had a Ia-CSM classification. These objects started interacting relatively soon after the explosion and remained active long enough to be picked up by our pipeline beyond the 100 day threshold. Figure~\ref{known_CSM_plots} shows the light curves of the recovered SNe Ia-CSM in absolute magnitude space (uncorrected for extinction). Even if the peak identified by our code is not the real peak due to it not being observed (e.g. for SN 2018evt, SN 2019agi, and SN 2019ibk), the CSM interaction persists for long enough for it to be picked up by our pipeline.

Ten of the known Ia-CSM SNe were presented in \citet{Ia-CSM_BTS}, who searched for SNe Ia-CSM discovered in the ZTF Bright Transient Survey from May 2018 to May 2021 \cite[BTS;][]{BTS-I, BTS-II}. They found two objects that are not in our sample: SN 2020abfe (ZTF20acqikeh) and SN 2020aekp (ZTF21aaabwzx). SN 2020abfe is in the DR2 sample, but due to a combination of a gap in the observed light curve and the interaction not altering the declining tail sufficiently, our tail fit procedure is unable to distinguish it from a normal declining SN Ia tail. SN 2020aekp was first detected after the final date for objects to be included into our sample. Two of the events in our sample (SN 2020eyj and SN 2020kre) were not presented in \citet{Ia-CSM_BTS}. SN 2020eyj was excluded as \citet{Ia-CSM_BTS} focused on interaction with H-rich material and this object showed He emission lines suggesting interaction with He-rich material \citep{Kool_He_CSM}. SN 2020kre was not in the BTS sample and therefore, not included in \cite{Ia-CSM_BTS}. However, it was confirmed with spectroscopy to have \Halpha~emission in its peak spectra.

Out of the 13 events SN 2020aeuh is an outlier, due to its distinct light curve. While the other 12 known Ia-CSM events detected in our sample have decline tails whose slopes are significantly shallower than for a normal SN Ia or steepen over time, SN 2020aeuh brightens significantly, having a double peaked nature with the second peak at around 100\,d after the first. Even though the light curve suggests the SN to be interacting, no H emission (\Halpha~or other lines) are observed. Kool et al.~(in prep) presents a detailed analysis of this object.

\subsubsection{Siblings}

\begin{figure*}
 \centering
 \includegraphics[width=16cm]{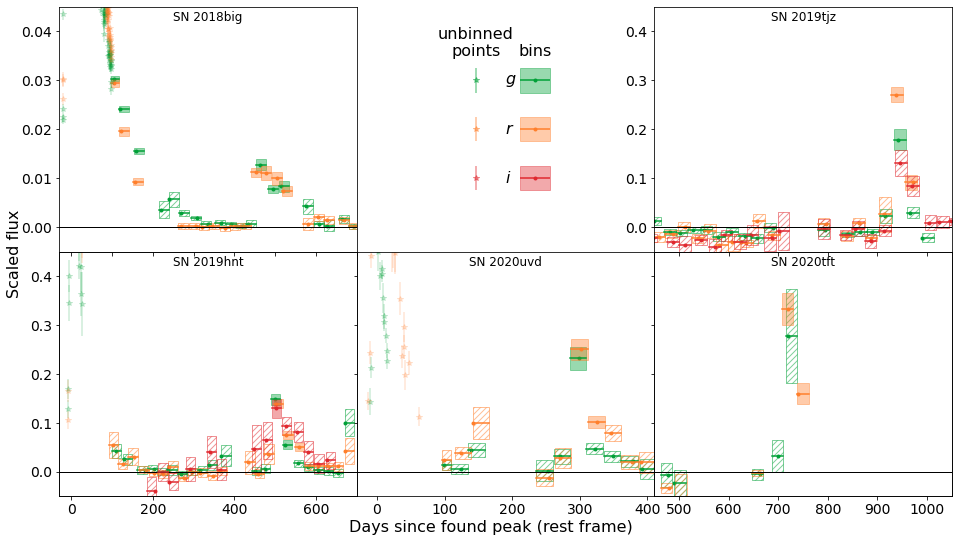}
 \caption{Binned late-time observations in flux space of the five events with a detected sibling, with the flux normalized to the found peak flux. All objects are plotted on the same flux scale for easy comparison except for SN 2018big, as its late-time detections are much weaker compared to the original SN peak magnitude due to the larger distance offset between the siblings.}
 \label{sibling_plots}
\end{figure*}

\begin{figure}
 \centering
 \includegraphics[width=\columnwidth]{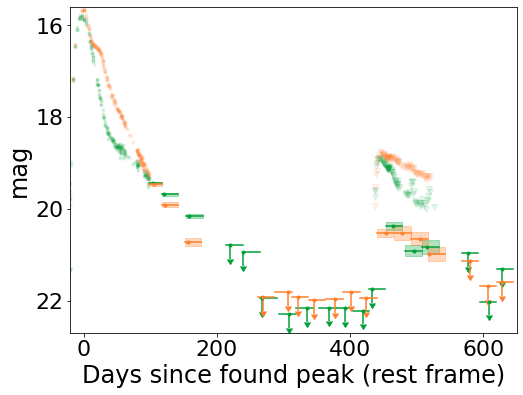}
 \caption{The light curves of SN 2018big, and its sibling SN 2019nvm in magnitude space using bins of 25 d. The \textit{g} (green) and \textit{r} (orange) bins follow the tail of SN 2018big until it disappears in the noise. About 450\,d after the peak of SN 2018big, new detections are identified in the binned photometry. The individual observations remain upper limits, although their shape hint to the true nature of these late-time detections.}
 \label{sibling_example}
\end{figure}

\begin{table*}
 \centering
 \caption{Objects with a detected sibling transient.}
 \begin{tabular}{llllcccc}
  \hline
  Primary name & IAU name & Sibling name & IAU name$^b$ & Type & Date$^a$ & Offset ($\arcsec$)$^c$\\
  \hline
  ZTF18aasdted & SN 2018big & ZTF19abqhobb & SN 2019nvm & IIP & 58\,714 & 3.7\\
  ZTF19aaysiwt & SN 2019hnt & ZTF20acwpads & - & - & 59\,186 & 1.2\\
  ZTF19acihfxz & SN 2019tjz & ZTF18aanhpii & - & - & 59\,774 & 1.2\\
  ZTF20abzetdf & SN 2020tft & - & - & Ia & 59\,867 & $<$ 1\\
  ZTF20acehyxd & SN 2020uvd & ZTF21abouuow & SN 2021udv & Ia & 59\,422 & 3.1\\
  \hline
 \end{tabular}
 \begin{flushleft}
$^a$MJD of the first detection of the sibling. \\
$^b$The siblings ZTF19aaysiwt and ZTF20acwpads share an IAU name. For ZTF20abzetdf, the siblings are too close together ($<$1$\arcsec$ separation) to be automatically recognised as separate events, causing them to share both ZTF and IAU names. ZTF18aanhpii was a sibling transient in 2022 on top of the host nucleus, resulting in the internal name being from 2018. \\
$^c$Angular separation on the sky between the siblings. \\
\end{flushleft} 
 \label{siblings}
\end{table*}

Siblings are transients that occur in the same host galaxy as each other and can be useful for understanding differences in local environments \citep[e.g.][]{biswas_siblings, ZTF_siblings}. In some cases, the siblings occur in (almost) the same place on the sky, only differing in explosion time. This can be either due to the two transients being physically close together, or a projection effect due to the inclination of the host. However, the result is the same: forced photometry at the location of one sibling will result in a (partial) recovery of the other. Assuming that the first transient is a SN Ia in our sample and the second transient is fainter, our pipeline will flag the late-time rebrightening as a late-time excess in one of our objects.

Careful examination of the images using \textsc{snap} and cross-referencing using Fritz (an alert broker, \citealt{skyportal2019, duev2019realbogus, Kasliwal_Growth, Skyportal}) and the Transient Name Server\footnote{https://www.wis-tns.org/} (TNS) showed that there are five objects in our shortlist whose late-time detections are due to a sibling. Figure~\ref{sibling_plots} shows the binned light curves of these objects in flux space. In each light curve there is a sudden significant spike in the detected fluxes in all observed bands, which falls back down again after a short period of time. Table \ref{siblings} lists the name and type of each sibling if known, as well as their sky separation. In some cases the siblings are close enough together that they are not automatically recognised as separate events, resulting in them having the same name. In the case of SN 2019tzj the sibling (ZTF18aanhpii) exploded close to the nucleus, which had some spurious detections in 2018. This caused the sibling to have a 2018 ZTF name, although it exploded in 2022.

Fig.~\ref{sibling_example} shows the detection of a sibling (SN 2019nvm) in the late-time light curve of SN 2018big in magnitude space. SN 2019nvm is slightly offset ($\sim4\arcsec$) from the location of the original SN. The photometry pipeline forces the point spread function (PSF) fit at the position of SN 2018big. As the position of SN 2019nvm is slightly offset, only some of the total flux of SN 2019vnm is captured in the fit. 
Besides these five siblings, we identified two other pairs of siblings with \textsc{snap}: SN 2019gcm and SN 2021fnj, and SN 2020jgs and SN 2021och. These siblings were too far apart to be picked up with the forced photometry (9.6 and 10.8$\arcsec$, respectively) but were found while inspecting using \textsc{snap}. For a complete list and study on the siblings found in the ZTF DR2, we refer the reader to Dhawan et al.~(in prep.).

\begin{figure*}
 \centering
 \includegraphics[width=16cm]{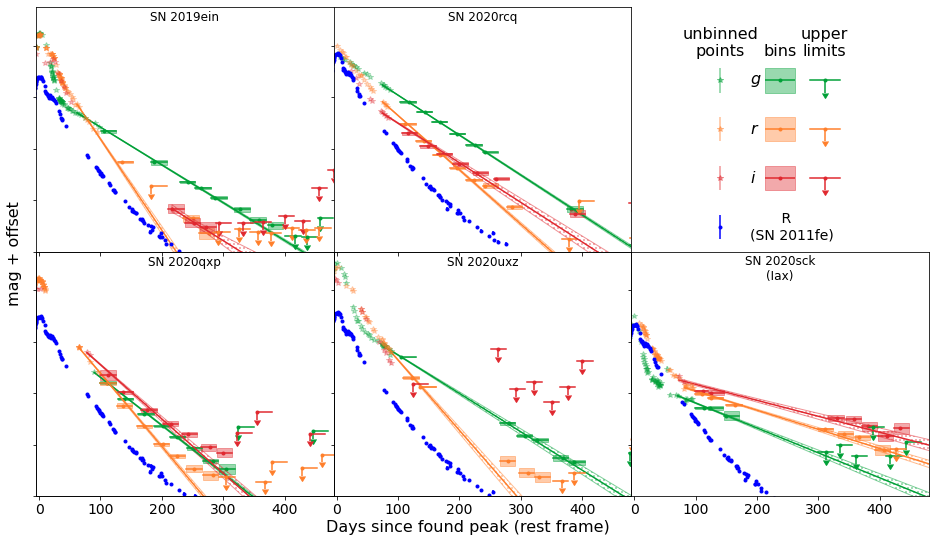}
 \caption{The five objects with kinks in their tails that start to deviate from the assumed decline rate at $\sim$200 -- 250\,d post peak are shown in magnitude space as a function of days since peak. As most of these objects have their peak magnitude blinded, no scaling is shown. A normal radioactive decay model has been fitted to these tails, shown as solid straight lines with their $1\sigma$ uncertainty as dashed regions. But as the ejecta opacity changes over time so does the half-life time of the tail, causing a kink seen in the bins which is not reproduced by the model. The arbitrarily normalised \textit{R}-band light curve of SN 2011fe (known not to have CSM interaction from detailed spectral studies) \citep{spec_Lijiang-2.4m} is shown in blue, showing the same shift in decline slope at a similar phase. }
 \label{kink_plots}
\end{figure*}

\subsubsection{Kinked tails}
\label{kink_tails}
This group consists of five objects where the simple tail model, based on the typical decline rate of SNe Ia based on SN 2011fe \citep{Georgios_11fe}, failed to fit the observations at later times. The reason for this failure in four of the events is that there is a slow-down in the $r$- and $i$-band decline rates at $\sim$200 -- 250\,d after peak, which the model does not take into account. The fifth event, SN 2020sck \citep{2020sck_Iax}, is a known SN Iax, a subclass known for having lower ejecta velocities and luminosities, suggesting that the explosion did not necessarily fully disrupt the star \citep{Iax_model_1, Iax_model_2}. This event was flagged because of a slow-down in its decline rate roughly 80 days after the identified peak. The change in slope is visible in all bands and significantly longer than the assumed $t_{1//2} = 50$\,d of normal SNe Ia. The presence of a bound remnant has been suggested to be the cause of similar late-time signatures seen in other SNe Iax \citep{Kawabata_iax14dt, close_Iax, Camacho-Neves_iax14dt}.

The four SNe objects that deviate from the simple tail model are very nearby ($z \leq 0.009$) compared to the majority of ZTF DR2 sample and when using the binned observations are bright enough to be detected up to (nearly) a year after their first detection. As Rigault et al. (in prep.) are using these nearby events to calibrate their H$_0$ measurement, their peak magnitudes are currently blinded. Figure~\ref{kink_plots} shows the \textit{g}-, \textit{r}-, and \textit{i}-band light curves of these objects in magnitude space. The \textit{R}-band light curve of SN 2011fe \citep{spec_Lijiang-2.4m}, which had a similar change in decline slope, is also shown for comparison. As discussed in \citet{Georgios_11fe}, the radioactive decays produce $\gamma$-rays and positrons, as well as X-rays and electrons that can be thermalised, depositing their energy in the expanding SN ejecta. As the ejecta expand over time they become more transparent, shortening the delay between thermalisation of the deposited energy and the emission of optical radiation. This results in the SN tail slope changing as the opacity changes.

To be able to observe a change in decline slope like this, a SN has to be both bright and well observed during the time it is visible. The other SNe Ia in our sample at similarly low redshifts have gaps in their observations or the change in slope is not strong enough for the tail fits to fall outside the allowed range of reduced chi-squared values ($\chi^2_\text{red} > 5$) and therefore, are not flagged by the pipeline due to this issue.

\subsection{Additional tests of promising events}
\label{Additional_tests}
 After performing the tests discussed in the previous sections on each event, there are 10 objects remaining with an unexplained late-time excess. Light echoes produced by SN light scattering off of nearby dust clouds were considered, but were ruled out as these are typically $\geq$10 mag fainter than the SN at its peak \citep{Patat_light_echoes, 2012cg}. We performed several additional tests on these 10 events to try to determine the origin of their late-time excess. The first is using SMP as described in Section \ref{lc_data}. The second is testing for coincidence with an AGN and third is more detailed comparisons with known transient classes. A summary of these tests is shown in Table \ref{alt_trans_res} and discussed below.

\begin{table*}
 \centering
 \caption{Results of the additional tests for the 10 promising objects. The host separation is given in \arcsec~and converted to kpc using the redshift given in Table~\ref{33_list} and the same cosmology as was used in Sec.~\ref{impact_refdepth}. In the final five columns the similarity of each SN Ia with a late-time excess is compared to different transient classes to see if the late-time signal can be interpreted as another transient. This can be excluded based on an inconsistent colour (1), duration (2), and/or an excessive amount of host extinction (3) needed to obtain the observed magnitudes.}
 \resizebox{\textwidth}{!}{
 \begin{tabular}{lccccccccccccc}
  \hline
   Name & Red.$^a$ & \multicolumn{2}{c}{Host separation} & \textit{E(B - V)}$_\text{host}$ & AGN$^c$ & Approx. & SN Ia? & Ib? & Ic? & IIP? & TDE? \\
   & error & (\arcsec) & (kpc) & (mag.)$^b$ & &mag.$^d$ & \\
   \hline
   SN 2018grt* &no&$0.36\pm0.03$ &$0.32\pm0.02$ & 0.21 -- 0.36 & no &$-$16.5& no (23) & no (23) & no (2) & no (3) & no (13)\\
   SN 2020pkj &no&$0.52\pm0.02$ & $0.28\pm0.01$ & 0.23 -- 0.36 & no &$-$15.4& yes & yes & yes & no (23) & no (13) &\\
   SN 2019ldf* &no&$0.65\pm0.04$ &$0.78\pm0.05$ & $\leq 0.03$ & no &$-$16.4& no (123) & no (123) & no (123) & no (123) & no (13) \\
   SN 2019mse &no&$0.59\pm0.05$ &$1.09\pm0.10$ & 0 & no &$-$17.5& no (2) & no (2) & no (2) & no (2) & yes &\\
   SN 2019rqn &no&$1.39\pm0.08$ &$2.21\pm0.13$ & $\leq 0.04$ & no &$-$16.8& no (13) & yes & yes & no (1) & no (13) \\
   SN 2019vzf &no&$3.89\pm0.08$ &$4.86\pm0.10$ & 0 & yes &-& - & - & - & - & - &\\
   SN 2020awr &yes&$19.60\pm0.04$ &$31.65\pm0.07$ & - &-& - & - & - & - & - & - &\\
   SN 2020alm &no &$0.67\pm0.07$ &$0.85\pm0.09$ & $\leq 0.05$ & no &$-$16.9& no (123) & no (12) & no (23) & no (123) & yes &\\
   SN 2020kzd &yes&$4.67\pm0.04$ &$8.07\pm0.07$ & - &-& - & - & - & - & - & - &\\
   SN 2020tfc* &no &$0.21\pm0.02$ &$0.14\pm0.01$ & $\leq 0.12$ & no &$-$16.8& no (123) & no (12) & no (23) & no (123) & no (3) &\\
   \hline
 \end{tabular}}
 \label{alt_trans_res}
\begin{flushleft}
$^a$Reduction error: Comparison between the standard light curve reduction and scene modelling identified issues with the baseline correction (see Section \ref{sec:scene_modelling}).\\
$^b$Estimated assuming the main peak is a normal SN Ia. When correcting for Milky Way extinction and distance is enough to exceed an absolute $g$-band magnitude of $-$19.3, we quote a host \textit{E(B -- V)} = 0 mag. \\
$^c$The presence of an AGN was estimated using the WISE colours of the host and the criteria of \cite{WISE_crit}.\\
$^d$Mean absolute magnitude of the \textit{r}-band late-time excess after correction for Galactic and host extinction, averaged over the host extinction range considered.\\
\end{flushleft}
\end{table*} 

\begin{figure*}
 \centering
 \includegraphics[width=\textwidth]{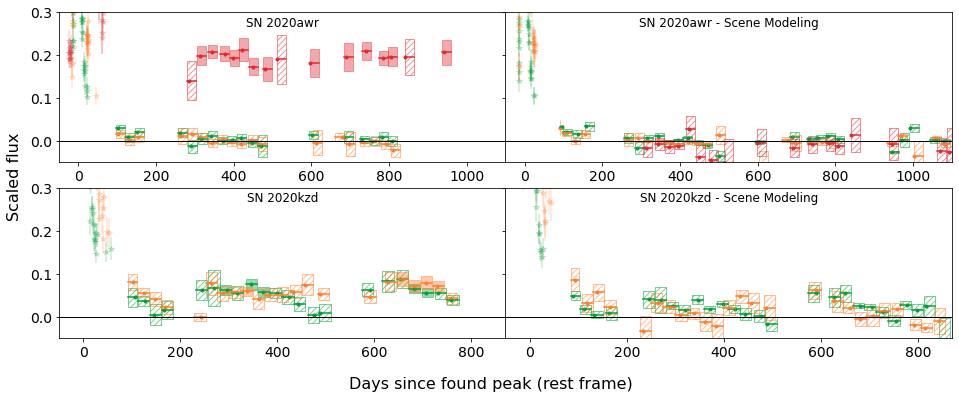}
 \caption{The two objects whose late-time detections were revealed to be caused by the photometry extraction. The colours are as in Fig.~\ref{kink_plots}. The left side shows the binned forced photometry light curve, and the right side shows the binned SMP light curve. Bins with $5\sigma$ detections are shaded solid, while the non-detections are hashed. }
 \label{other_red_errs}
\end{figure*}

\subsubsection{Scene modelling photometry}
\label{sec:scene_modelling}
To test for issues in data processing, we ran the 10 events through the scene modelling pipeline of Lacroix et al.~(in prep.) as discussed in Section \ref{lc_data}. Two SNe Ia (SN 2020awr and SN 2020kzd) were found to have issues with their difference imaging and forced photometry light curves, causing false detections. The detections in both these events are close to the detection limit and therefore, are impacted by even small errors in the baseline placement. The difference imaging (left panel) and scene-modelling (right panel) light curves of these two events are shown in Fig.~\ref{other_red_errs}. 

For SN 2020awr, $\sim$300\,d after the SN peak, the \textit{i}-band observations jump up to detections at $\sim$20.3 mag. In contrast to other objects where baseline issues were found, here the offset occurs only for part of the light curve, while the pre-SN baseline has no visible issues. We could not identify a clear reason for this when inspecting the images with \textsc{snap}, and the SN is too far away from the host nucleus for it to be host activity. For the scene-modelling version of the photometry, the jump in the \textit{i}-band observations has disappeared completely, showing that there was indeed an unidentified issue with the \textit{i}-band data for this object. Since the late-time detections are determined to be spurious, this object is ruled out from having late-time detections.

In the case of SN 2020kzd, late-time detections are present in the \textit{g}- and \textit{r}-bands for hundreds of days (see Fig.~\ref{other_red_errs}). The SN is in a complex environment with three galaxies close to its sky position, which likely complicates the image subtraction and baseline correction. When SMP is performed on the event the detections disappear and average flux at late times is consistent with zero, showing that the binned forced photometry light curve likely suffered from a wrongly determined baseline correction.

\subsubsection{AGN contamination}
\label{sec:agn_cont}
If a SN explosion site is coincident with a host galaxy that has an AGN, host activity is a likely cause of the late-time detections. Using data from the Wide-field Infrared Survey Explorer (WISE; \citealt{WISE}), \citet{WISE_crit} present a criterion to test if a galaxy hosts an AGN based on the WISE \textit{W1--W2} and \textit{W2--W3} colours, which we apply to our events. One object (SN 2019vzf) is 4.86 kpc of its host centre and its host is a known AGN, with WISE colours of \textit{W1--W2} = $0.55\pm0.03$ and \textit{W2--W3} = $2.90\pm0.04$ mag. In \textsc{snap}, the late-time signal appears to cover both the SN and AGN locations. The AGN contamination is too strong to put any meaningful constraints on the late-time flux at the SN location. We kept the object in our sample until now to test if it was possible to use scene modelling to reduce the AGN contamination. However, this was not possible. We attribute the late-time signal to host activity and disregard it in future discussion. The light curves of SN 2019vzf are shown in Fig. \ref{other_alt}.

\begin{table}
 \centering
 \caption{
 Details of the comparison transients used to test if late-time detections could be explained by another transient at a similar sky position. The first column shows the assumed type of transient, the second shows the transients used to represent each type, the third has the approximate absolute extinction-corrected $r$-band peak magnitudes, and the fourth the reference for each event. }
 \resizebox{\columnwidth}{!}{
 \begin{tabular}{lccc}
  \hline
   Type & Name & Peak abs. M$_{r}$ (mag.) & Reference\\
   \hline
   SN Ia & SN 2011fe & $-$18.4 -- $-$19.4 & \citet{spec_HST}\\
   SN Ib & SN 2019yvr & $-$17.9 & \citet{Ib_ex}\\
   SN Ic & SN 2021krf & $-$17.3 & \citet{21krf_ext}\\
   SN IIP & SN 2020jfo & $-$17.8 & \citet{IIp_ext}\\
   SN IIP & SN 2017gmr & $-$18.7 & \citet{2017gmr}\\
   TDE & AT 2018hco & $-$22.1 & \citet{TDE_ext}\\
   TDE & AT 2018zr & $-$20.1 & \citet{TDE_ext}\\
   \hline
 \end{tabular}}
 \label{alt_trans}
\end{table} 

\subsubsection{Presence of a sibling close to the SN location}
To test if a previously unidentified sibling transient is causing the late-time detections of the remaining seven objects, we have compared their late-time light curves to known classes of transients, including a SN Ia, core-collapse SNe (Type Ib, Type Ic, Type IIP) and two tidal disruption events (TDE). Firstly, we have estimated the amount of potential host extinction from the main SN peak by assuming it was a normal SN Ia with a typical $g$-band peak of $-18.8$ to $-19.3$ mag after correcting for the distance to the SN and for Milky Way extinction. These estimated host extinction values are given in Table \ref{alt_trans_res}, assuming R$_\text{V} = 3.1$. The bright end of the absolute peak magnitude gives an upper limit for the host extinction, and the faint end gives a lower limit. After correcting for this range of additional host galaxy extinction, the late-time excesses have mean absolute \textit{r}-band magnitudes of $-$15.4 to $-$17.5 mag (see Table \ref{alt_trans_res}). 

After estimating the allowed extinction for each primary SN Ia, we have initially assumed that if the late-time excess is due to another transient then it will have the same extinction along the line-of-sight. For these other transients, we used examples of a SN Ia, Ib, Ic, two IIPs, and two TDEs to compare against, with details of the comparison objects described in Table \ref{alt_trans}. The transients chosen to represent their category are all in the typical magnitude range for their type. Two TDE and SN IIPs were chosen to represent the upper and lower end of the range of peak magnitudes expected for these transients. We use our model for SN 2011fe to represent SNe Ia and test the lower and upper edge of the range of normal SN Ia peak $r$-band magnitude.

These comparison objects were chosen as they have well-sampled light curves in the ZTF filters, and literature values for the host extinction. We corrected the light curves of the comparison objects using their literature redshifts and their extinction values before correcting for the redshift and extinction of each SN Ia in our sample with a potential late-time excess. We then compare this transient light curve to the found late-time detections to see how well they match. A good match would have a similar magnitude, colour, and duration. 

It could be the case that the suspected sibling was in the same line-of-sight direction, but had a different amount of extinction due to e.g. exploding behind a cloud that adds additional extinction. We check this by adding enough extinction to match the $r$-band detections between the different comparison events and the observed late-time detection and again check if the colour and duration match up, as the observed colour is affected by the extinction. For the TDE comparisons, we allow a host galaxy \textit{E(B -- V)} of up to one magnitude, as was estimated for the ZTF TDE sample of \citet{TDE_host_ext_range}.

\begin{figure*}
 \centering
 \includegraphics[width=\textwidth]{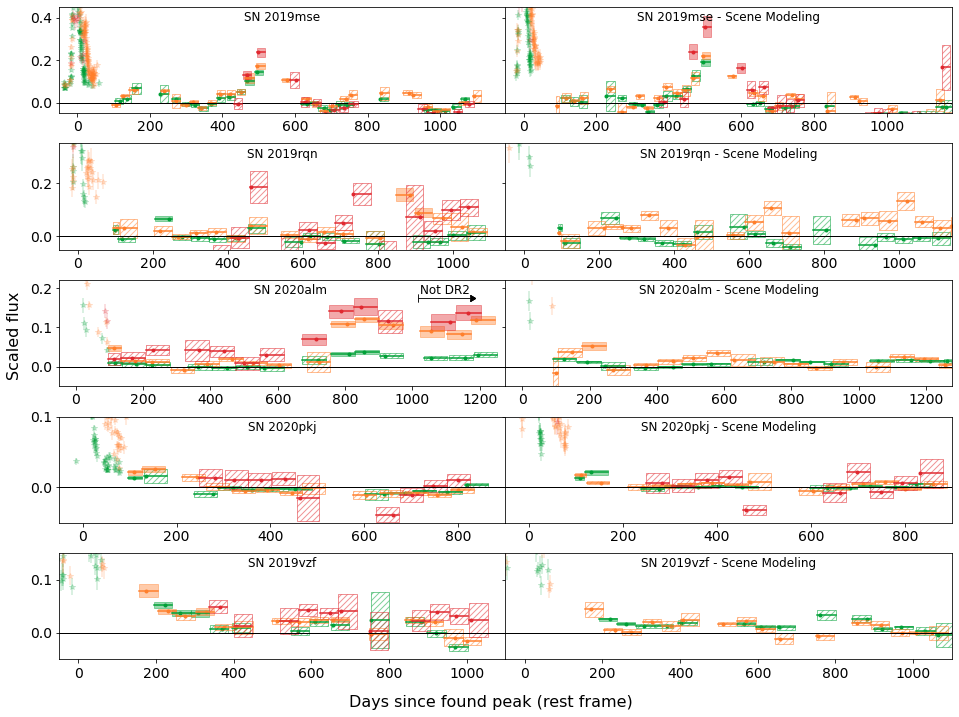}
 \caption{The top four rows show the light curves of the objects where a previously undetected sibling transient as an explanation for the late-time observations could not be ruled out, with forced photometry light curves in the left-hand panels and the scene modelling photometry light curves in the right-hand panels. The $5\sigma$ detections are shown as bins with solid uncertainty regions and bins with hashed uncertainty regions are non-detections. The object whose late-time detections are caused by the host galaxy AGN, SN 2019vzf, is shown in the bottom row. The colours are as in Fig.~\ref{kink_plots} with \textit{g} band in green, \textit{r} band in orange, and \textit{i} band in red. }
 \label{other_alt}
\end{figure*}

For four events (SN 2019mse, SN 2019rqn, SN 2020alm, and SN 2020pkj), the late-time detections are consistent with at least one of the comparison classes, as detailed in Table \ref{alt_trans_res} and shown in Fig.~\ref{other_alt}. We describe them individually in the following sections. The three remaining events (SN 2018grt, SN 2019ldf, and SN 2020tfc) can not be explained by the presence of a sibling transient and are discussed further in Section \ref{sec:late_time_cand}.

\subsubsection*{SN 2019mse}
SN 2019mse has late-time detections starting at 450\,d post peak and lasting $\sim$250\,d in the \textit{gri}-bands with an absolute \textit{r}-band magnitude during the excess of $-$17.5 mag. Careful re-examination of the difference images showed that the late detections are slightly offset (about one pixel) from the SN location, and appear to be on the host nucleus location instead. However, with its WISE colours being \textit{W1 -- W2} = $0.25\pm0.04$ and \textit{W2 -- W3} = $2.44\pm0.09$, the host is determined to not contain an AGN. The scene modelling version of the light curve shows a similar late-time excess, showing that this is not an artifact from the chosen photometric analysis method. 

The late-time signal is detected in all three bands, and its behaviour is very similar in all of them (see Fig.~\ref{other_alt}). Its rise and decline time scales, absolute magnitude, and colours agrees well with the ranges seen for TDE. Together with the observation that the late-time detection are at the host nucleus location, this suggests that a nuclear transient explains the late-time signal adequately.

\subsubsection*{SN 2019rqn}
In the case of SN 2019rqn, there is a short period of detections at 950 -- 1050\,d in the $r$-band after a gap in the observations, declining and fading below the detection significance within 100\,d of the first detection. Nothing is detected at a $\geq 5 \sigma$ level in the $g$ or $i$-band observations. The \textit{i}-band SMP light curve had its host contribution not completely subtracted, causing a flux offset in the data points. We therefore do not consider the SMP \textit{i}-band further. The data points in the \textit{g}- and \textit{r}-band light curve have slightly larger uncertainties in the SMP version, causing the main SN light curve to fall below the 5$\sigma$ threshold at an earlier epoch resulting in fewer individual points visible (Fig.~\ref{other_alt}). Similarly, larger uncertainties for the SMP prevents the detection of a late-time \textit{r}-band signal.

Assuming a sibling exploded during the gap in the observations between 870 and 950\,d post peak, SNe Ib and Ic with \textit{E(B -- V)} $\leq 0.3$ mag extinction in the $r$-band can fit their tail to match the observed detections in the $r$-band without being excluded by the $g$-band and $i$-band non-detections. Therefore, we cannot rule out a sibling as the source of the detected late-time signal.

\subsubsection*{SN 2020alm}
The late-time signal in SN 2020alm is seen in all three bands, beginning at $\sim$750\,d post peak and lasting at least 300 d. There is a gap of 80\,d in the observations immediately before the period of activity. The detections slowly rise to a plateau. Our initial analysis only included data for SN 2020alm up to $\sim$1000\,d after peak. However, when this object was identified as having late-time detections, the light curve pipeline was rerun and it was found to be still bright at later times, with significant detections in the $r$- and $i$-bands, but not above 5$\sigma$ detections in the $g$-band. The \textit{i}-band SMP light curve contained a significant flux offset, as the host was not fully subtracted. We therefore do not further consider this band. The binned SMP \textit{r}-band light curve fails to reproduce the late-time detections found in the forced photometry light curve. However, the \textit{g}-band detections are recovered in the SMP. The SN is close to the host nucleus at 
 0.66\arcsec~(0.85 kpc) offset at the redshift of the SN, but the WISE colours of \textit{W1--W2} = $0.06\pm0.04$ and \textit{W2--W3} = $2.42\pm0.12$ place it far outside the AGN region. 

As SN 2020alm was still active while our analysis was on-going, we obtained two spectra using the Optical System for Imaging and low-Intermediate-Resolution Integrated Spectroscopy (OSIRIS) instrument on the Gran Telescopio CANARIAS (GTC) at Roque de los Muchachos in La Palma on 26 July 2023 using the R1000R grism. As the spectrum is heavily dominated by the host galaxy, we subtracted a rebinned spectrum of the host taken by the Sloan Digital Sky Survey (SDSS, \citealt{SDSS-I-II, SDSS_DR4, SDSS_telescope, SDSS_Spectograph}) in 2003, well before the SN occurred. We confirmed successful host subtraction by checking for residual \NaID~and \MgI~${\lambda5175}$ absorption lines and found that no residual features were present. A more detailed explanation is given in Appendix \ref{spec_sec}.

The resulting spectrum shows an excess that is stronger towards longer wavelengths (Fig.~\ref{ZTF20aaifyfx_spec}). This is consistent with the broadband photometry finding an brighter excess in the redder bands, while the $g$-band remains within the noise after binning the observations. There is some excess in the narrow [\NII]~${\lambda\lambda6548,6583}$, \Halpha~and [\SII]~$ {\lambda\lambda6716, 6730}$ emission lines, but we lack the resolution to check if this is significant. There is no visible additional \Halpha~component in the spectrum, which would be indicative of CSM interaction. Integrating the spectrum over the $r$- and $i$-band efficiencies gives an $r-i$ colour of 0.6 mag, which is within $3\sigma$ of the value found in the latest photometry bin.

\citet{TDE_host_ext_range} showed that TDEs generally have a $g$ -- $r$ colour of zero and can have featureless spectra. We approximate a TDE by a flat line in order to estimate the amount of extinction needed to generate a red excess similar to the spectrum. We find that the general shape of the spectrum can be approximated with $0.6 < \text{E(B -- V)}_\text{host} < 1$ mag. This would mean an absolute $r$-band magnitude of $-18.8 > M_r > -19.8$ mag. \citet{TDE_host_ext_range} showed that both this amount of host extinction and late-time brightness are possible for TDEs. They also showed that it is possible for a TDE to rise and fall back down within the 80\,d gap in observations, although this is seen in fainter TDEs than corresponding to our estimated absolute magnitude range.

The TDE sample of \citet{TDE_host_ext_range} did not contain a single object that matches our late-time detections in duration and luminosity in SN 2020alm. However, it is possible to combine parts of different TDEs together to make a TDE that peaked and decayed within the 80\,d gap and levelled out by the time it became observable again. Based on this, a TDE is a plausible explanation for the late-time signal detected in this object.

\subsubsection*{SN 2020pkj}
In the case of SN 20120pkj, the first \textit{r}-band bin with a detection is the end of the normally declining tail, but after the first bin the detections rise slightly in the next \textit{r}-band bin (Fig.~\ref{other_alt}). None of the binned photometry for the \textit{g}- and \textit{i}-bands give significant detections. Unfortunately, there is a gap in the observations immediately after the \textit{r}-band detections preventing us from following its evolution closely at these phases. When it became observable again at $>$200 d, no significant detections were found in any band. The duration of the transient is at least 75\,d and could be up to 150 d. In the SMP light curve of this object, a similar rise at the same epochs is recovered, though it is found in the \textit{g}-band instead of the \textit{r}-band. This is likely due to the low significance of these detections at just 5.8-$\sigma$ in the \textit{r}-band in the forced photometry and 5.2-$\sigma$ in the \textit{g}-band in the SMP, showing that they are just on the detection limit of our binning technique.

With a redshift of $z = 0.02489$, this object is nearby enough for the end of the tail to be visible in the bins at 100 days. This could explain why the first bin is a detection, but the slight increase is still unexpected for a normal SN Ia decline tail. While the second bin is marginally consistent with our tail fit, a slight but steady increase can be seen in the unbinned flux values as well, suggesting that the brightening is real.

Assuming that the detected part of the late-time signal was the brightest part of a sibling transient, the tested transients need significantly more extinction of $\sim$ 3 to 6 mag in the \textit{r}-band (depending on the comparison transient) than was found for the main SN Ia peak. However, the $\sim 80$\,d gap is long enough for a Type I sibling SN to peak higher during the gap and dim again before observations resumed. The other transient could have just started when it stopped being observable, and declined below the detection limit when the location became observable again. Since there is no constraint on the magnitude, this could work with any amount of extinction. Therefore, we cannot rule out conclusively that the late-time detections are due to another transient.

\subsubsection{Late-time interaction candidates}
\label{sec:late_time_cand}
\begin{figure*}
 \centering
 \includegraphics[width=\textwidth]{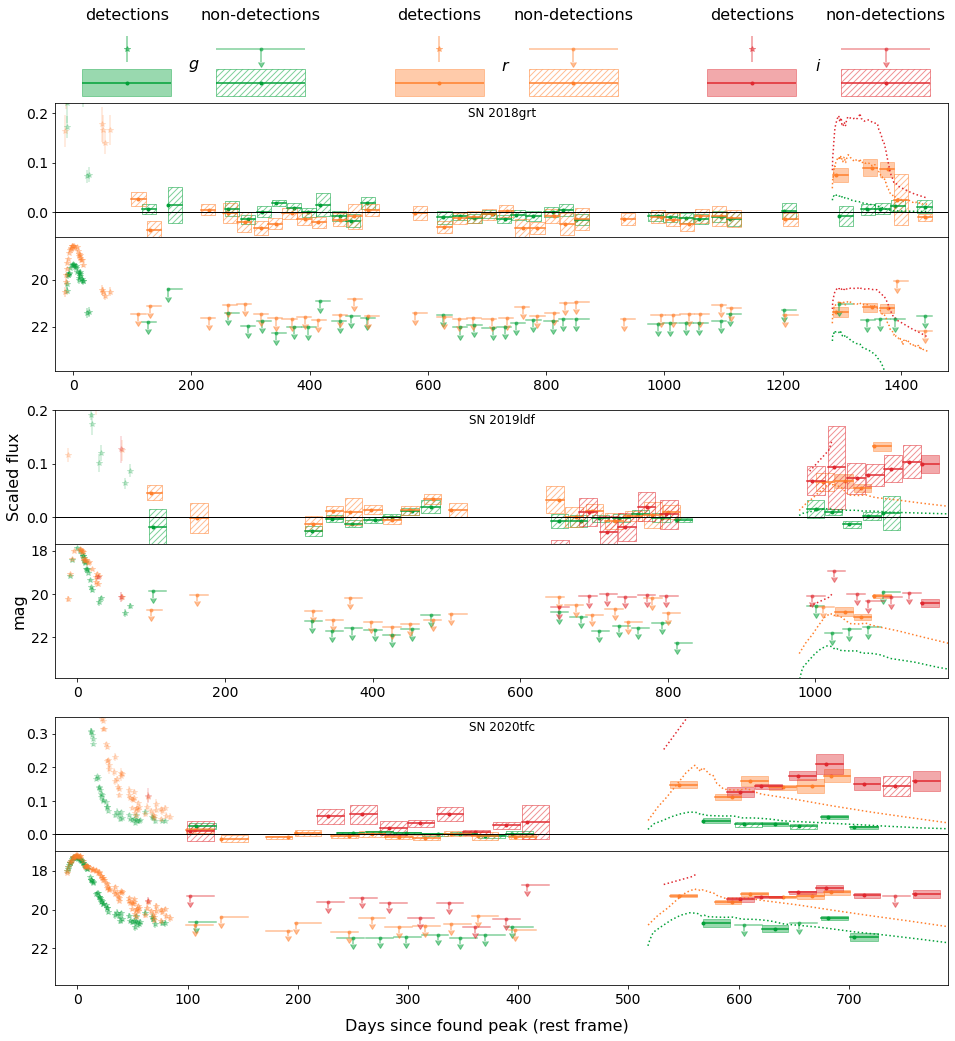}
 \caption{The three candidate objects, shown in magnitude and flux space. All three have significant detections ($\geq5\sigma$) after a period of observations consistent with zero flux. From the alternative explanations the best fitting alternate transients are shown in dotted lines. For SN 2018grt this is the Type IIP SN 2017gmr, for SN 2019ldf and SN 2020tfc this is the TDE AT 2018hco.}
 \label{candidates}
\end{figure*}

\begin{figure*}
 \centering
 \includegraphics[width=13cm]{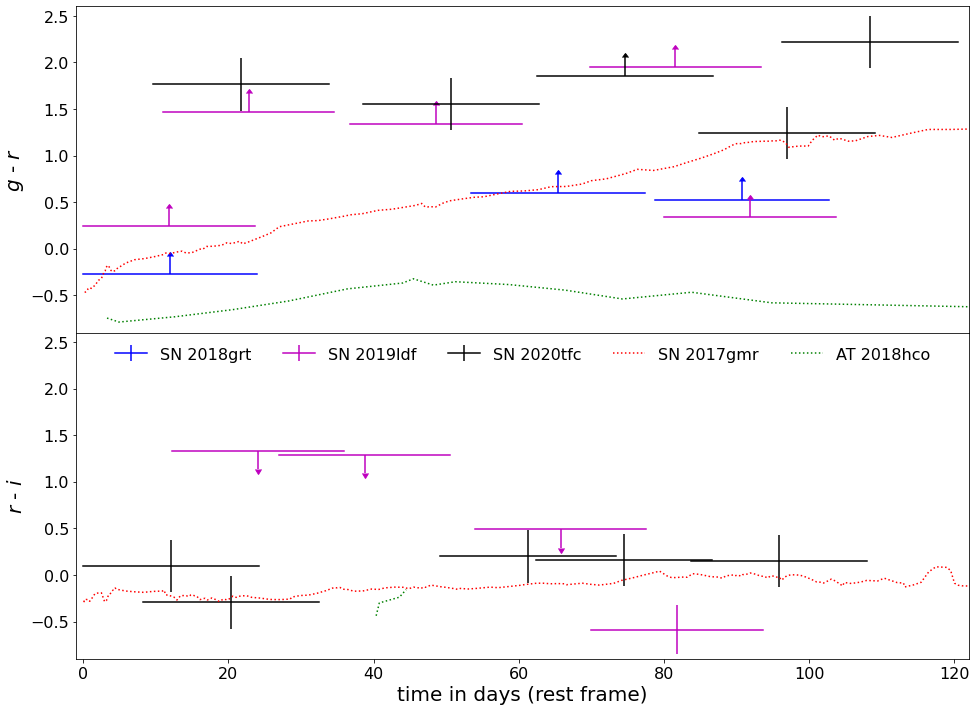}
 \caption{$g$-$r$ and $r$-$i$ colour curves of the three candidate objects, together with the colours of the best fitting alternate transients. The first bin for each object starts at zero days, but the bins can be shifted horizontally in an attempt to better fit the colour curve of the transient compared against (given that this is allowed by the rest of the light curve). Bins whose mean observation dates are closest to each other are used to calculate the colour, provided that these bins overlap in time. If there is a detection in only one band used to calculate the colour while the other is a non-detection, the result is a lower or upper limit.}
 \label{candidate_colours}
\end{figure*}

Finally, we present the three objects (SN 2018grt, SN 2019ldf, SN 2020tfc) whose late-time detections could not be explained by any of the explanations discussed above. Their light curves are shown in Fig.~\ref{candidates}, and the colours of the late-time detections in Fig.~\ref{candidate_colours} and the SMP light curves for SNe 2019ldf and 2020tfc in~\ref{final_candid_SMP}. A scene modelling analysis could not be performed for SN 2018grt due to it exploding early in the ZTF survey. The forced photometry pipeline uses images obtained during the ZTF commissioning phase as templates but the SMP does not. We present these three objects as late-time interaction \textit{candidates} as there are no spectra of these objects at these late times to confirm the presence of features consistent with CSM interaction.

By using the estimated mean absolute $r$-band magnitude after removing all extinction effects (see Table \ref{alt_trans_res}) we can estimate the required \Halpha~flux assuming that it is the source of all the detected flux in the $r$-band, and that the emission line has the same width as was used during the simulations. The identified strength of the signal is compared to the \Halpha~signal detected in SN 2015cp for each of the events.

\subsubsection*{SN 2018grt}
The late-time signal of SN 2018grt is only detected in the \textit{r}-band, the \textit{g}-band stays around zero flux, and there are no \textit{i}-band observations at these phases. The first detection in the \textit{r}-band begins at 1350\,d post peak with a magnitude of $21.4\pm0.2$ (absolute magnitude of $-$16.4 mag). It varies little over the $\sim 100$\,d period where it is detected, after which it returns to zero flux within 50\,d. The SN is close to the host nucleus with an offset of 0.35\arcsec~(0.32 kpc) but its host colours place it well outside the expected AGN region. Checking the difference images with \textsc{snap} shows that the host nucleus and SN location differ by $\sim$1 pixel. 

The late-time detections are $\sim 2.5$ mag below the main SN peak. If the late-time signal is due to another SN Ia or a TDE, it would require a significantly higher extinction value than was found for SN 2018grt itself. In addition, this object shows a sudden drop in the $r$-band, which is not normal behaviour for a TDE. Most of the other transient types cannot reproduce the plateau followed by the sharp decline only detectable in the $r$-band, apart from a SN IIP, where the plateau of SN 2017gmr has nearly the same time span. However, to fit the observed magnitude with a IIP SN, it would require a host $ E(B$ -- $V)$ that is three times higher than what was found for SN 2018grt, making this scenario less likely to be the case. Therefore, we conclude that late-time CSM interaction is a plausible scenario for the late-time signal in this object.

If we assume $0.21 \leq E(B$ -- $V)_\text{host} \leq 0.36$ mag and that the $r$-band signal is produced only by an \Halpha~emission line with a similar width to the one used in our simulations in Section \ref{simulation}, we estimate the strength of the emission to be much stronger than SN 2015cp, at 60 to 100 times its emission strength. However, there are examples of SNe Ia-CSM with interaction strengths this strong, e.g. SN 2020eum was within this range \citep{Ia-CSM_BTS}.

\subsubsection*{SN 2019ldf}
SN 2019ldf has late-time detections in the \textit{r}-band beginning at 1050\,d post peak and lasting for about 100 d, with an additional increase in brightness towards the end. There is a single $5\sigma$ detection in the \textit{i}-band but there are a number of lower significance detections coeval with the \textit{r}-band detections. These detections are directly after a long period without observations due to the object being behind the Sun. Nothing is detected in the \textit{g}-band during the time of the rise in the \textit{r} band. The binned SMP light curve recovers these late-time \textit{r}-band detections, and a single \textit{i}-band detection, showing that these detections are not specific to the photometry method.

We have compared the properties of the late-time detections to those of our comparison transient objects. Even if we assume that a SN exploded during the gap in observations in order to avoid needing a significantly larger \textit{E(B -- V)}$_\text{host}$ value, SNe evolve too much over a period of 100 days to explain the detections. In addition, detections would also be expected in the $g$-band, which have not been found. 

A TDE could fit the detections if it was intrinsically bright but heavily extincted, as this could explain the red colour and absence of signal in the $g$-band. However, SN 2019ldf is offset from the host nucleus by 0.65 \arcsec~(0.78 kpc) and inspection of the difference images using \textsc{snap}, shows that the late-time signal is more consistent with the SN location than the host nucleus location, disfavouring the TDE explanation. Therefore, we conclude that the late-time signal could be due to late-time CSM interaction.

The late-time detections persist until the end of the observation window. To determine if there were still signs of interaction once it was visible again, we obtained $g$- and $r$-band photometry with the EFOSC2 imaging spectrograph \citep{EFOSC2} on the ESO New Technology Telescope (NTT) in La Silla, Chile on 2023 May 19 as part of the extended Public ESO Spectroscopic Survey of Transient Objects+ (ePESSTO+; \citealt{PESSTO}).

To examine whether the SN is still detected in our images from May 2023, we used image subtraction techniques. Due to the lack of reference images in the $g$- and $r$-band filters from EFOSC2, we used images from the DESI Legacy Imaging Surveys Data Release 9 \citep{DESI-Legacy_Imaging_Surveys}. After aligning the images, we subtracted them from each other with the High Order Transform of Psf ANd Template Subtraction code version 5.11 \citep[\textsc{hotpants};][]{HOTPANTS}. We measured the brightness in the difference images using aperture photometry. The photometry was calibrated against stars from DESI Legacy Imaging Surveys. The EFOSC2 and DESI Legacy Imaging Surveys filters are not identical which might add an unknown systematic to the reported photometry. The $5\sigma$ upper limits are $m_g = 24.7$ and $m_r = 24.3$ mag at 1397 days after the peak, with no detection in either band. This means that the signal has disappeared at this time, and thus could have lasted, at most, for about 500 days.

Assuming that the $r$-band signal is entirely due to the \Halpha~emission, we estimate it to be $\sim$60 times as strong as the late-time interaction found in SN 2015cp. However, this assumption is very simplistic, as it completely disregards the rise in the $i$-band and therefore, it is only a first order estimate.

\subsubsection*{SN 2020tfc}
This object has late-time \textit{gri}-band detections, beginning at 550\,d post peak and lasting for at least 250 days. While the $r$- and $i$-bands are at more or less the same magnitude, the $g$-band detections are about 1.3 mag fainter. This immediately poses an issue for any alternate transient considered, as either there is a low amount of extinction and a weak $g$-band signal, or there is a high amount of extinction and the intrinsic signal is even brighter in the $i$-band. This, combined with the fact that the signal lasts for several hundreds of days with little variation, disfavours a SN as an alternate transient explanation. The intrinsic colour also heavily disfavours a TDE as these objects tend to have a similar intrinsic brightness in the $g$-, $r$-, and $i$-bands. In the binned SMP light curve the late-time detections were confirmed in the \textit{g}-band. The SMP \textit{i}-band data points have a large scatter, most likely due to uncertain background removal because of a low number of available images for the SMP template. Therefore, we do not consider them further in our analysis. In the \textit{r}-band SMP light curve, the data points are below our 5-$\sigma$ cut-off for detections, although one is very close to our limit at 4.6$\sigma$. However, the \textit{g}-band detections seen in the forced photometry are confirmed by SMP suggesting a real signal is present at late-times in at least this band, and at lower significance in the \textit{r} band. 

Similar to SN 2020alm, the late-time signal was on-going during our analysis but unfortunately, there is no archival host galaxy spectrum available to compare to. As the host dominates the late-time signal (the SN is at a distance of 0.28\arcsec (0.21 kpc) from the host nucleus) and cannot be removed using difference imaging or SMP like for the photometry, this prevented us from taking a spectrum of the late-time signal. With all alternate explanations ruled out or severely challenged by observations, the late-time CSM interaction remains as a plausible explanation for these late-time detections. 

If we assume that the $r$-band signal is due to \Halpha~emission only, the interaction is estimated to be 110 to 150 times as strong as the interaction found in SN 2015cp. This is by far the strongest of the three, but again this simple assumption is unrealistic as it completely ignores the measured $g$- and $i$-band signal that suggests a contribution from a continuum or other spectral lines to the late-time signal. 
 
\begin{table*}
 \centering
 \caption{The parameters used for rate estimation simulations for each object. For each of the three SNe we assumed a worst and best case scenario for detecting them in our sample, giving upper and lower limits for a simulated observing campaign. The efficiency curves generated from these simulations were then used to determine the intrinsic fraction of SNe Ia with late-time CSM interaction in an MCMC process, with the assumption that only one object was recovered in a sample size that is the same as the amount of DR2 objects with observations after the start epoch ($N_\text{sample}$). The last two columns show the found late-time CSM interaction fraction of the total SN Ia rate and the late-time CSM interaction rate.}
 \begin{tabular}{lcccccc}
  \hline
  Object & Start epoch & Duration & Strength & $N_\text{sample}$ & late-time CSM & late-time CSM\\
   & (d) & (d) & (SN 2015cp) & & fraction & rate (Gpc$^{-3}$ yr$^{-1}$)\\
  \hline
  SN 2018grt (worst) & 1\,375* & 100 & 60 & 748 & $0.0084_{-0.0041}^{+0.0183}$ & $203_{-97}^{+438}$\\[4pt]
  SN 2018grt (best) & 1\,275* & 200 & 100 & 988 & $0.0015_{-0.0007}^{+0.0032}$ & $36_{-18}^{+76}$\\[4pt]
  SN 2019ldf (worst) & 1\,050* & 200 & 60 & 1\,931 & $0.0019_{-0.0009}^{+0.0036}$ & $45_{-22}^{+87}$\\[4pt]
  SN 2019ldf (best) & 875* & 500 & 60 & 2\,505 & $0.0023_{-0.0011}^{+0.0038}$ & $54_{-26}^{+91}$\\[4pt]
  SN 2020tfc (worst) & 550 & 250 & 100 & 3\,439 & $0.0004_{-0.0002}^{+0.0009}$ & $10_{-5}^{+22}$\\[4pt]
  SN 2020tfc (best) & 450 & 500 & 150 & 3\,493 & $0.0003_{-0.0002}^{+0.0008}$ & $8_{-4}^{+20}$\\
  \hline
 \end{tabular}
\begin{flushleft}
$*$When the start epoch of the late-time excess was $>$750 d, 750\,d was used as the start epoch because of limitations of the available ZTF survey plan. See Section \ref{rates_csm} for more details.\\
\end{flushleft} 
 \label{rate_sims}
\end{table*}

\subsection{Late-time CSM interaction rates based on our candidate objects}
\label{rates_csm}

 We have identified three objects (SNe 2018grt, 2019ldf, and 2020tfc) with potential late-time CSM interaction signatures. For SN 2018grt and SN 2019ldf, these detections were in the \textit{r}-band (with no \textit{i}-band data available for SN 2018grt and low significance \textit{i}-band detections for SN 2019dlf). For SN 2020tfc, late-time detections were found in all three bands (\textit{gri}). In our initial simulations to determine our recovery efficiency (Section \ref{simulation}), we made the assumption that any CSM interaction would be dominated by \Halpha~emission that would be present in only the \textit{r}-band up to \textit{z} $ \sim$ 0.07 and in the \textit{i}-band beyond that. This was based on the dominant interaction signatures seen in SNe Ia-CSM and also the one event with late onset interaction, SN 2015cp \citep{2015cp}. 

In \citet{2015cp}, \CaII\ NIR triplet emission was also identified in its spectra, along with emission consistent with \MgI~$\lambda5175$. \cite{2018ApJ...868...21H} speculated that although the \CaII\ NIR region was very noisy, the \CaII\ emission may be a similar strength to the \Halpha~emission. For SN 2019ldf at a redshift of 0.057, the \CaII\ NIR triplet, if present, would be partially shifted out of the \textit{i}-band, which could potentially explain its low-significance \textit{i}-band detections. SN 2020tfc is at lower redshift (0.031) so the \CaII\ NIR triplet would fall completely in the \textit{i}-band and could potentially be the source of the \textit{i}-band detections. The strength of the potential \MgI~$\lambda5175$ line in SN 2015cp was weak and not well constrained \citep{2015cp} but could potentially result in a weak signature in the \textit{g}-band. However, without spectral confirmation this is only speculation.

The other difference between our initial simulations for determining recovery efficiency and the observed detections in our three candidate late-time CSM objects is that the late-time CSM interaction signature is much stronger in our observed events. In our three candidates, the late-time signal is between 2 to 3 magnitudes below the SN peak magnitude, while in the simulations with strongest interaction the signal was about 4.4 magnitudes below the peak.

It is clear from our candidate events, that there is significant diversity (detected bands, timescales and strengths) in their potential CSM signatures and without knowledge of their underlying spectra (e.g. emission lines that are present), it is difficult to develop spectral models and run simulations covering this full diversity. Therefore, although very simplistic and ignoring other potential emission lines occurring in the \textit{g}- and \textit{i}-bands as were seen in SN 2015cp \citep{2015cp}, we have focused on \Halpha~emission dominated models and the corresponding \textit{r}-band detection efficiency to estimate a rate of late-time \Halpha~emission dominated CSM interaction based on our three candidate objects. We use the same underlying spectral model (SN 2011fe combined with \Halpha~emission) as detailed Section \ref{simulation} in further \textsc{simsurvey} simulations but constrain the strength and timescale of the CSM signature based on our three events. In Table \ref{rate_sims}, we show the range of start epoch, duration and strength (dependent on assumed host galaxy extinction) simulated for each event. For each of the three objects, we take the worst case scenario (shortest, weakest, latest start epoch) and the best case scenario (longest, strongest, earliest start epoch) allowed by the constraints from the data to estimate a worst and best case scenario recovery efficiency for each object.

\begin{figure*}
 \centering
 \includegraphics[width=\textwidth]{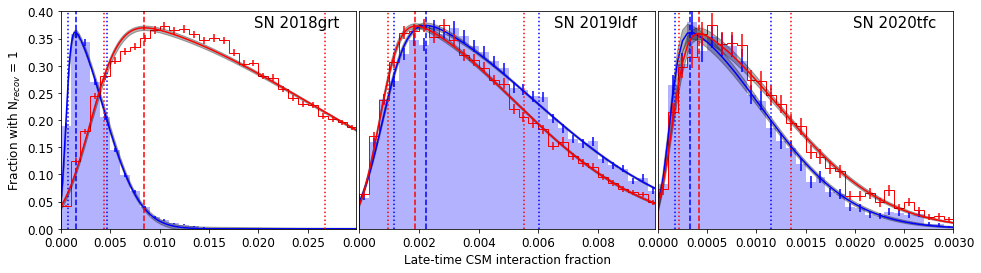}
 \caption{Fraction of MCMC realisations per bin that resulted in one object being recovered in a sample with the same size as the effective DR2 sample size as a function of the late-time CSM interaction fraction. The best and worst case scenario for each object is shown in blue and red, respectively. A skewed normal distribution fit with a $1 \sigma$ uncertainty band is shown for each scenario, and the dashed and dotted lines give maximum and 68 per cent confidence interval of these distributions, respectively. The distributions continue on the right side of each plot.}
 \label{rate_fig}
\end{figure*}

Since the available ZTF survey plan for our simulations only spans 1004 d, we cannot set the interaction signal to start at phases after this as the interaction would never be observed. Therefore, we choose to start the interaction in our simulations no more than 750\,d after the peak. This is done to ensure there is a decent chance to observe an event until after the interaction has started, while still being able to apply a good baseline correction. As the interaction is by far the dominant source of light at these epochs, changing the start date of the interaction like this has little effect on the total brightness of the SN. However, it does affect the number of objects observed by allowing more objects to have observations at later times, when the interaction occurs. This is a redshift independent effect, but will impact the resulting recovery fraction function. To account for this, we only consider objects with observations after the start of the interaction, and remove those where the simulation limits prevent any possible late-time detection. While changing the interaction start date has increased the number of objects that satisfy this condition, there are still some objects that do not due to e.g. being in a sparsely observed field.

Not all objects in the DR2 have detections up to the time after the peak where the interaction was detected in the three final candidates. Similar to the simulation, in order to get a proper estimation of the rate we can only consider the objects that could have been found interacting ($N_\text{sample}$), and remove those without observations this late after the peak. This gives us an effective sample size that is smaller than the full DR2, and is listed in Table \ref{rate_sims}.

The expected number of observed objects with late-time interaction in a sample $N_\text{recov}(N_\text{sample}, \eta(z))$ is a function of the efficiency and the intrinsic rate, which in turn is a fraction of the total SN Ia rate. If we assume the SN Ia rate from \citet{SNIa_rate} as in our initial simulations and that the fraction of these showing late-time interaction, $f$, is redshift independent, then the only $z$ dependency is in the efficiency of our pipeline. In a similar approach to \citet{SLSN_rate} and \citet{Ca-rich_rate}, we run a Markov chain Monte Carlo (MCMC) simulation with $10^7$ realizations for each scenario to find the fraction of interacting SNe that best explains our findings.

In each realization of the MCMC, we draw $N_\text{sample}$ objects, and assign a redshift to each object according to the distribution from \citet{SNIa_rate}. For each object we also draw two random numbers from a flat distribution between zero and one. The first random number is used to decide if the object was interacting, and the second to decide if interaction would have been recovered. Interaction is true if the random number is below the chosen value for $f$ for that realization. Recovery is true if the second random number is below $\eta(z)$ at the $z$ of the object (e.g. if the recovery efficiency is 0.7 at the SN redshift, a drawn number below 0.7 will result in a detection if the object was interacting). The number of recovered objects with interaction can be found by counting the objects for which both interaction and recovery is true, while the actual number of interacting objects can be found by counting the objects where interaction is true.

Figure \ref{rate_fig} shows per bin the fraction of realisations that resulted in one object being recovered as a function of late-time CSM interaction fraction with the best and worst scenario parameters for the three discovered events. We approximate them by fitting a skewed normal distribution and use the fit to estimate the peak of the distribution and the 68 per cent confidence interval on either side. These values are quoted as a fraction of the total SN Ia rate and as the late-time CSM interaction rate in the last two columns of Table \ref{rate_sims}.

\section{Discussion}
\label{discussion}

In this section, we first discuss the overall rate of CSM interaction signatures found in our analysis (Section \ref{discuss_interaction}) and some of the properties of the three SNe Ia displaying these unexplained late-time detections (Section \ref{discuss_prop}). 

\subsection{Late-time interaction is rare in SNe Ia}
\label{discuss_interaction}

In our study of the $\sim$3500 SNe Ia in the ZTF DR2 with photometry at $>$100 d, we have identified three objects (SN 2018grt, SN 2019ldf, and SN 2020tfc) for which late-time CSM interaction is the best explanation for their detected late-time flux excesses. However, as each of these had significantly different parameters for the interaction, they had to be treated separately when trying to estimate the rate of late-time CSM interaction in SNe Ia (Section \ref{rates_csm}). Except for the worst case scenario of SN 2018grt, the identified rate for all cases agree with one another within the uncertainty, with the fraction of SNe Ia showing late-time \Halpha-dominated CSM interaction between $0.03^{+0.08}_{-0.02}$ and $0.23^{+0.38}_{-0.11}$ per cent. This is similar to what was found by \citet{Ia-CSM_BTS} in their Ia-CSM sample, and agrees with the upper limit set by \citet{2015cp} with their discovery of the late-time interaction in SN 2015cp. Some objects were disregarded as we could not definitively determine that they were not due to a sibling transient. Therefore, there could be a handful of additional objects displaying strong late-time \Halpha-dominated CSM interaction that were not included in our rate estimates but this would double the rate at most. Therefore, we estimate that the intrinsic rate of late-time \Halpha-dominated CSM interaction in SNe Ia is $<0.5$ per cent.

The identified late-time signals that are consistent with CSM interaction occurred between 1.5 to over 3.5 years after the original SN peaked. If we assume that the SN ejecta have a velocity of the order of $10^4$ km s$^{-1}$, the distance at which the CSM shell resides is of the order of $10^{17}$ cm. As CSM moves slowly, this means it must have been ejected from the system a long time ago. The further out the shell is from the progenitor system, the more mass is contained in even a thin shell. This is especially true when one considers that in order to get a detectable signal from the interaction, the CSM density cannot be too low. At these distances, light travel time effects also become significant, and a short interaction with a thin shell will be smeared out over a long time. Unfortunately, since we have only partial constraints on the interaction timescales and CSM mass estimation from an interaction signature is far from straightforward, we do not attempt to provide a CSM mass estimation.  

An interesting aspect of the late-time detections in our three candidates is their inferred strength. While our model is likely an oversimplification by trying to explain the entire $r$-band signal using only a narrow \Halpha~line, it is clear that the signal is much stronger compared to the one found in SN 2015cp, especially when there are other band(s) at a similar magnitude. Interaction this strong is not unprecedented though, SN 2020aekp for instance, starts to plateau around 50 days after peak at an absolute magnitude M$\sim-18.5$ in all three ZTF bands \citep{Ia-CSM_BTS} and holds this plateau for several hundreds of days. Most of the known SNe Ia-CSM presented in \citet{Ia-CSM_BTS} have detections in multiple bands, and the spectra presented show more than just an \Halpha~line coming from the CSM interaction. CSM around a SN Ia also does not have to be spherically symmetric, especially in a scenario where material has been stripped and partially lost from the donor star in the progenitor system. \citet{IIn_asym_CSM} show for SNe IIn that if the CSM mostly resides in a disk, only the ejecta travelling in the direction of the disk will be slowed down and interact while ejecta at higher inclinations continues to expand normally. 

The colour curves for the late-time detections in our three objects (Fig.~\ref{candidate_colours}), show that $g - r \sim$1.7 mag for SN 2020tfc, with limits of $g - r > 0.5$ and 1.5 mag for SN 2018grt and SN 19ldf, respectively.
For the two objects with $i$-band observations, $r-i<0.5$ mag. The SN Ia-CSM sample of \cite{Ia-CSM_BTS} had late-time ($>$300 d) $g - r$ colours of $0 - 0.5$ mag, which are similar to SN 2018grt but bluer than SN 2019ldf and SN 2020tfc. \citet{Kool_He_CSM} find $g-r\sim-1$ mag and $r-i\sim0.5$ mag at late times for the He-interacting Ia-CSM SN 2020eyj. When taking into account the host galaxy extinction that they identify, it becomes even bluer, which is much bluer than our three objects.

\subsection{Properties of SNe Ia with late-time interaction}
\label{discuss_prop}

The three SNe Ia with late-time flux excesses that can not be explained by other scenarios have SALT2 light curve fitter $x_1$ and $c$ values that are generally typical of normal SNe Ia (Rigualt et al.~in prep.). However, the $c$ value of SN 2018grt of 0.61 $\pm$ 0.03 is at the high end and would be excluded from cosmological analyses. We also identified significantly more host extinction for this object than for the other two based on their luminosities at peak. 

All three candidates were found at small projected distances from their host nuclei. While the small sample size might be used to explain part of this observation, the mean projected distance of all objects in the ZTF DR2 is 6.3 kpc (Rigault et al.~in prep.). The chance for our three objects to be at most 0.8 kpc from the host is $<1$ per cent, suggesting that objects with late-time interaction have a preference for small host separations, the detections are caused by nuclear variability, \textit{or} they are caused by bad host subtraction. The location of the late-time signal is too close to the host nucleus to distinguish which location the late-time signal is more consistent with by using \textsc{snap}. The host galaxy colours are inconsistent with being AGN and the properties of the late-time detections cannot be easily explained by known nuclear transient classes, such as TDEs. It is possible that by going to deeper limiting magnitudes with our light curve binning that we have identified previously unstudied nuclear variability but this cannot be confirmed. Imperfect template subtraction can occur to galaxy centres, resulting in a dipole artifact. However, these are easily recognisable with \textsc{snap}, and this possibility was ruled out for our three candidates. Therefore, we conclude that that a likely interpretation is that there may be something intrinsic to SNe Ia exploding in these environments that produces late-time CSM interaction but further samples are required to confirm. 

The morphologies of the host galaxies of the candidate events are broadly elliptical/spheroidal in nature with no signs of spiral arms. The galaxies have stellar masses of 4 $\times$ 10$^9$, 4 $\times$ 10$^9$, and 2 $\times$ 10$^{10}$ M$_\odot$ for SN 2018grt, SN 2019ldf, and SN 2020tfc, respectively. These host masses are all within the bulk of the masses of the ZTF DR2 SN Ia sample (Rigault et al.~in prep.). The hosts of the known H-rich Ia-CSM sample have stellar masses of $\sim$3 $\times$10$^8$ to 3 $\times$ 10$^{10}$ M$_\odot$, consistent with our objects \citep{Ia-CSM_BTS}. SN 2020eyj, the He-interacting and radio detected SN Ia-CSM \citep[SN 2020eyj;][]{Kool_He_CSM} also had a very small offset from its host of $0.57\pm0.02\arcsec$ ($0.36\pm0.01$ kpc). Its host only has a WISE \textit{W3} upper limit so it can not be excluded from having an AGN present. It is a compact star forming galaxy with a mass of $\sim$6 $\times$ 10$^7$ M$_\odot$, which is lower than the typical Ia-CSM range \citep{Ia-CSM_BTS}. 

The hosts of the three candidate events all have WISE \textit{W1 -- W2} $\approx0$ mag. SN 2020tfc has \textit{W2 -- W3} of $\sim$0.8 mag, while the other two have only limits with \textit{W2 -- W3} $>$ 1.2 mag. Based on Fig.~2 of \cite{Irani_wise}, this places them in the overlap region between elliptical and spiral host properties from the galaxy sample of \cite{Lintott_galaxyzoo}. The vast majority of Ia-CSM sample of \cite{Ia-CSM_BTS} has \textit{W2 -- W3} colours of $>$ 1 mag, again broadly consistent with our sample. However, the morphologies of our candidate events do not show evidence for spiral arms and their overlap between ellipticals and spirals in the \textit{W2 -- W3} parameter space suggest they may come from different, older stellar populations than the known Ia-CSM events \citep{Kool_He_CSM, Ia-CSM_BTS}. However, to prove that these are different populations requires a larger sample. 

\subsection{Limitations of the analysis}
Since there is no good model of a Type Ia SN interacting with CSM at late times that could be used for our \textsc{simsurvey} simulations, we had to make our own based on several assumptions. Our main assumptions were that the SN Ia looks normal until the moment the interaction starts, and we assumed that this interaction was with hydrogen-rich material showing itself primarily as an \Halpha~emission line. This is motivated by the fact that \Halpha~is found in most SNe Ia-CSM and the late-time interaction in SN 2015cp was confirmed through the observation of an \Halpha~emission line. However, \citet{2015cp} also found other emission lines that are associated with the SN, like the \CaII~triplet near 8500 \AA~and a tentative detection of \MgI~$\lambda5175$, suggesting that our model is too simplistic. Similarly, for SN 2020tfc, we identified late-time detections in the \textit{gri}-bands, which cannot be explained by \Halpha~emission alone. 

This method of binning the late-time light curves allows to push the detection limit down to a limiting magnitude m = 21.44 mag. Since the references used in ZTF have a mean limiting magnitude of $\overline{\text{m}}_\text{lim} \sim 21.8$ mag, this becomes the leading uncertainty preventing our binning technique to go deeper. We have shown through our simulations that deeper references will allow for the binning technique to go deeper and recover fainter interaction signatures in a larger redshift volume. We have also used SMP to generate light curves that did not rely on forced photometry to test for issues coming from the photometry measurement. In most cases the late-time detections were recovered with both methods, showing their robustness. However, the SMP has been found to have issues with identifying a baseline flux when the numbers of images used in scene model template creation is small. This highlights the requirement for a long `off' to model the underlying galaxy light sufficiently. 

The pipeline focuses more on the historical light curve of the SNe than their current state. Since we require two adjacent bins to be detections in order to avoid false positives and the smallest bins we use are 25 d, the interaction needs to be active for over a month before it is picked up. Only after this, can it be followed up with spectroscopy if identified fast enough and the interaction signature is still occurring. Unfortunately, in the cases of our three candidates this could not be done due to a combination of recovering the interaction late, the objects being blocked by the Sun for a section of the year, and not having a suitable reference spectrum of the host galaxy to subtract and isolate the interaction spectrum. Adapting the pipeline to run in real-time would help with shortening the timescale to detection but if the host galaxy contribution is strong, then an archival reference spectrum is required.

\subsection{By-product pipeline detections}
The conservative, catch-all approach outlined in our work poses the challenge of identifying exactly why each object was flagged by the pipeline. Our approach allows for different kinds of objects that become interesting at late times to be caught. We recovered nearly all known Ia-CSM SNe through their non-normal SN Ia tails, as well as a number of known and unknown siblings, along with evidence of a change in the declining tail slope of close-by, normal Type Ia SNe around 200 days after the peak. In our sample, four per cent of the objects needed to be visually inspected, upon which the cause of the (false) positive became clear quickly for most objects. Some of the checks, like using the WISE colours to identify an AGN host close to the SN location, have been automated, reducing the workload in future attempts to use this method for searching for late-time signals, especially when using large datasets like the Vera C.~Rubin Observatory's Legacy Survey of Space and Time \cite[LSST;][]{LSST}.

The pipeline, as presented in this paper, is specifically tailored for finding SNe Ia that deviate from normal behaviour at $>$100\,d after peak. However, the method of binning late-time observations can be used for any type of transient, but the checks to ensure that the resulting detections are not due to expected behaviour are SN Ia specific. When other types of transients (e.g. SN Ib, Ic, IIP) are used as input, the result will likely vary. For instance, a long lasting plateau in an SN IIP will be unexpected by the pipeline as it does not follow the decline tail of a normal SN Ia. However, modifying the pipeline to work for different kinds of transients mostly requires revision of these checks.

\section{Conclusions}
\label{conclusion}
We have presented a search of the ZTF DR2 SN Ia light curve sample for signatures of late-time CSM interaction. This is the first systematic search for signatures of late-time interaction in SNe Ia in a large optical survey. We made a custom pipeline to calibrate and bin the light curve data at more than 100\,d after the peak, using bins with sizes between 25 and 100 d. Our analysis was based on searching for \Halpha~emission (as was seen in SN 2015cp and Ia-CSM) in the \textit{r}-band at lower redshift and \textit{i}-band above a redshift of $z = 0.07$. We performed simulations with \textsc{simsurvey} to determine the efficiency of our search, as well as the intrinsic rate of potential \Halpha-dominated CSM interaction in the sample. Our main conclusions are:

\begin{enumerate}
 \item Our pipeline returned 134 SNe that were potentially interesting based on their late-time light curves. Visual inspection of these objects was performed using our visualisation program, \textsc{snap}, to inspect the difference images, removing 101 objects as false positives. 
 \item Of the remaining 33 objects, we identified 13 out of the 14 known Ia-CSM objects in the DR2, five siblings close to the position of the original SN Ia, four very nearby events whose late-time light curves were not captured by our simple radioactive tail model and one Iax with a late-time excess in all three bands, consistent with the presence of a bound remnant. 
 \item Out of our final shortlist of 10 candidate events, we identified three SNe Ia (SN 2018grt, SN 2019ldf, and SN 2020tfc) that displayed late-time detections beginning 550 - 1350\,d after peak and lasting at least 100 - 250\,d, which could not be explained by data issues, AGN activity, or other transient events exploding at a similar location. 
 \item For SN 2018grt, these late-time detections were only in the \textit{r}-band (no coeval \textit{i}-band data was available). For SN 2019ldf, detections were made in the \textit{ri}-bands and for SN 2020tfc in all three bands suggesting potential contributions with \CaII\ NIR emission or other \MgI\ as was identified in SN 2015cp \citep{2015cp}. 
 \item The \textit{r}-band magnitudes of the late-time interaction are $-16.5$, $-16.4$, and $-16.8$ mag for SN 2018grt, SN 2019ldf, and SN 2020tfc, respectively. At their respective redshifts, this corresponds to \Halpha~interaction strengths of 60 -- 150 times that of SN 2015cp (depending on the extinction correction used). The strong nature of this signal could suggest we might only have found the high end of the late-time interaction strength distribution.
 \item Using \textsc{simsurvey} simulations of the ZTF survey, we estimated the intrinsic rate of strong \Halpha-dominated late-time ($>100$ days after the SN peak) interaction to be occurring in $<$0.5 per cent of SNe Ia. This translates to absolute rates of $8_{-4}^{+20}$ to $54_{-26}^{+91}$ Gpc$^{-3}$ yr$^{-1}$, assuming a constant SN Ia rate of $2.4\times10^{-5}$ Mpc$^{-3}$ yr$^{-1}$ for $z \leq 0.1$.
\end{enumerate}

The rarity of late-time interaction (occurring in $<$0.5 per cent of SNe Ia) highlights the importance of a large dataset of objects that have been observed for multiple years. The late-time detections occurred at different epochs for each object (from 550 -- 1350\,d post peak), showing that the phase at which SNe Ia will start to show signs of CSM interaction is highly variable. Therefore, the only viable strategy is to keep observing SNe even after they have faded beyond the detection limit and binning the late-time light curves. The interaction strength that we are sensitive to is heavily dependent on both the science and reference image depth. Future improvements to the analysis would include the use of deeper reference images to detect fainter signatures of late-time interaction, as well as running the pipeline in real time so that additional photometry and spectroscopy can be obtained to further characterise the late-time excesses. The deeper magnitude limits of LSST would be ideal for this study but cannot be immediately performed when the survey starts because of the requirements of deep reference images, as well as the need to wait up to more than three years post SN Ia peak for the interaction to occur. 

\begin{acknowledgements}
JHT, KM and MD acknowledge support from EU H2020 ERC grant no. 758638. 
L.G. acknowledges financial support from the Spanish Ministerio de Ciencia e Innovaci\'on (MCIN), the Agencia Estatal de Investigaci\'on (AEI) 10.13039/501100011033, and the European Social Fund (ESF) “Investing in your future” under the 2019 Ram\'on y Cajal program RYC2019-027683-I and the PID2020-115253GA-I00 HOSTFLOWS project, from Centro Superior de Investigaciones Cient\'ificas (CSIC) under the PIE project 20215AT016, and the program Unidad de Excelencia Mar\'ia de Maeztu CEX2020-001058-M.
This work was funded by ANID, Millennium Science Initiative, ICN12\_009.
T.E.M.B. acknowledges financial support from the Spanish Ministerio de Ciencia e Innovaci\'on (MCIN), the Agencia Estatal de Investigaci\'on (AEI) 10.13039/501100011033, and the 
European Union Next Generation EU/PRTR funds under the 2021 Juan de la Cierva program FJC2021-047124-I and the PID2020-115253GA-I00 HOSTFLOWS project, from Centro Superior de Investigaciones Cient\'ificas (CSIC) under the PIE project 20215AT016, and the program Unidad de Excelencia Mar\'ia de Maeztu CEX2020-001058-M.
MN is supported by the European Research Council (ERC) under the European Union’s Horizon 2020 research and innovation programme (grant agreement No.~948381) and by UK Space Agency Grant No.~ST/Y000692/1.
Y.-L.K. has received funding from the Science and Technology Facilities Council [grant number ST/V000713/1].
TWC acknowledges the Yushan Young Fellow Program by the Ministry of Education, Taiwan for the financial support.

Based on observations obtained with the Samuel Oschin Telescope 48-inch and the 60-inch Telescope at the Palomar Observatory as part of the Zwicky Transient Facility project. ZTF is supported by the National Science Foundation under Grants No. AST-1440341 and AST-2034437 and a collaboration including current partners Caltech, IPAC, the Weizmann Institute of Science, the Oskar Klein Center at Stockholm University, the University of Maryland, Deutsches Elektronen-Synchrotron and Humboldt University, the TANGO Consortium of Taiwan, the University of Wisconsin at Milwaukee, Trinity College Dublin, Lawrence Livermore National Laboratories, IN2P3, University of Warwick, Ruhr University Bochum, Northwestern University and former partners the University of Washington, Los Alamos National Laboratories, and Lawrence Berkeley National Laboratories. Operations are conducted by COO, IPAC, and UW.

This work was supported by the GROWTH project \citep{Kasliwal2019} funded by the National Science Foundation under Grant No 1545949.

The Gordon and Betty Moore Foundation, through both the Data-Driven Investigator Program and a dedicated grant, provided critical funding for SkyPortal.

The Legacy Surveys consist of three individual and complementary projects: the Dark Energy Camera Legacy Survey (DECaLS; Proposal ID \#2014B-0404; PIs: David Schlegel and Arjun Dey), the Beijing-Arizona Sky Survey (BASS; NOAO Prop. ID \#2015A-0801; PIs: Zhou Xu and Xiaohui Fan), and the Mayall z-band Legacy Survey (MzLS; Prop. ID \#2016A-0453; PI: Arjun Dey). DECaLS, BASS and MzLS together include data obtained, respectively, at the Blanco telescope, Cerro Tololo Inter-American Observatory, NSF’s NOIRLab; the Bok telescope, Steward Observatory, University of Arizona; and the Mayall telescope, Kitt Peak National Observatory, NOIRLab. Pipeline processing and analyses of the data were supported by NOIRLab and the Lawrence Berkeley National Laboratory (LBNL). The Legacy Surveys project is honored to be permitted to conduct astronomical research on Iolkam Du’ag (Kitt Peak), a mountain with particular significance to the Tohono O’odham Nation.

Funding for the SDSS and SDSS-II has been provided by the Alfred P. Sloan Foundation, the Participating Institutions, the National Science Foundation, the U.S. Department of Energy, the National Aeronautics and Space Administration, the Japanese Monbukagakusho, the Max Planck Society, and the Higher Education Funding Council for England. The SDSS Web Site is http://www.sdss.org/.

The SDSS is managed by the Astrophysical Research Consortium for the Participating Institutions. The Participating Institutions are the American Museum of Natural History, Astrophysical Institute Potsdam, University of Basel, University of Cambridge, Case Western Reserve University, University of Chicago, Drexel University, Fermilab, the Institute for Advanced Study, the Japan Participation Group, Johns Hopkins University, the Joint Institute for Nuclear Astrophysics, the Kavli Institute for Particle Astrophysics and Cosmology, the Korean Scientist Group, the Chinese Academy of Sciences (LAMOST), Los Alamos National Laboratory, the Max-Planck-Institute for Astronomy (MPIA), the Max-Planck-Institute for Astrophysics (MPA), New Mexico State University, Ohio State University, University of Pittsburgh, University of Portsmouth, Princeton University, the United States Naval Observatory, and the University of Washington.

Based on observations made with the Gran Telescopio Canarias (GTC), installed at the Spanish Observatorio del Roque de los Muchachos of the Instituto de Astrofísica de Canarias, on the island of La Palma. This work is (partly) based on data obtained with the instrument OSIRIS, built by a Consortium led by the Instituto de Astrofísica de Canarias in collaboration with the Instituto de Astronomía of the Universidad Autónoma de México. OSIRIS was funded by GRANTECAN and the National Plan of Astronomy and Astrophysics of the Spanish Government.

The ztfquery code was funded by the European Research Council (ERC) under the European Union's Horizon 2020 research and innovation programme (grant agreement n°759194 - USNAC, PI: Rigault).

This work made use of Astropy:\footnote{http://www.astropy.org} a community-developed core Python package and an ecosystem of tools and resources for astronomy \citep{astropy:2013, astropy:2018, astropy:2022}.

Many people have contributed to lmfit. The attribution of credit in a project such as this is difficult to get perfect, and there are no doubt important contributions that are missing or under-represented here. Please consider this file as part of the code and documentation that may have bugs that need fixing.

Some of the largest and most important contributions (in approximate order of size of the contribution to the existing code) are from:
 Matthew Newville wrote the original version and maintains the project.
 
 Renee Otten wrote the brute force method, implemented the basin-hopping and AMPGO global solvers, implemented uncertainty calculations for scalar minimizers and has greatly improved the code, testing, and documentation and overall project.
 
 Till Stensitzki wrote the improved estimates of confidence intervals, and contributed many tests, bug fixes, and documentation.
 
 A. R. J. Nelson added differential\_evolution, emcee, and greatly improved the code, docstrings, and overall project.
 
 Antonino Ingargiola wrote much of the high level Model code and has provided many bug fixes and improvements.
 
 Daniel B. Allan wrote much of the original version of the high level Model code, and many improvements to the testing and documentation.
 
 Austen Fox fixed many of the built-in model functions and improved the testing and documentation of these.
 Michal Rawlik added plotting capabilities for Models.

 The method used for placing bounds on parameters was derived from the clear description in the MINUIT documentation, and adapted from J. J. Helmus's Python implementation in leastsqbounds.py.

 E. O. Le Bigot wrote the uncertainties package, a version of which was used by lmfit for many years, and is now an external dependency.

 The original AMPGO code came from Andrea Gavana and was adopted for lmfit.

 The propagation of parameter uncertainties to uncertainties in a Model was adapted from the excellent description at https://www.astro.rug.nl/software/kapteyn/kmpfittutorial.html\#confidence-and-prediction-intervals, which references the original work of: J. Wolberg, Data Analysis Using the Method of Least Squares, 2006, Springer.

Additional patches, bug fixes, and suggestions have come from Faustin Carter, Christoph Deil, Francois Boulogne, Thomas Caswell, Colin Brosseau, nmearl, Gustavo Pasquevich, Clemens Prescher, LiCode, Ben Gamari, Yoav Roam, Alexander Stark, Alexandre Beelen, Andrey Aristov, Nicholas Zobrist, Ethan Welty, Julius Zimmermann, Mark Dean, Arun Persaud, Ray Osborn, @lneuhaus, Marcel Stimberg, Yoshiera Huang, Leon Foks, Sebastian Weigand, Florian LB, Michael Hudson-Doyle, Ruben Verweij, @jedzill4, @spalato, Jens Hedegaard Nielsen, Martin Majli, Kristian Meyer, @azelcer, Ivan Usov, and many others.

The lmfit code obviously depends on, and owes a very large debt to the code in scipy.optimize. Several discussions on the SciPy-user and lmfit mailing lists have also led to improvements in this code.

Other software: numpy \citep{numpy}, pandas \citep{pandas_software, pandas_paper}, matplotlib \citep{matplotlib}, sncosmo \citep{sncosmo}, simsurvey \citep{simsurvey}

\end{acknowledgements}

\section*{Data Availability}
The ZTF DR2 photometry is available on GitHub. The binning program is available at \url{https://github.com/JTerwel/late-time_lc_binner}. \textsc{snap} is available at \url{https://github.com/JTerwel/SuperNova_Animation_Program}. The ePESSTO+ photometry is available on the ESO archive. The late-time spectrum of SN 2020alm will be available upon request to the author.

%
\bibliographystyle{aa}
\bibliography{main}

\begin{appendix}
\onecolumn
\section{Tables}
\begin{longtable}{cccccc}
 \caption{Spectra used to make the SN 2011fe model. All spectra were taken from WISeREP \citep{wiserep}.}
 \label{11fe_sources}
 \endfirsthead
 \hline
 MJD  & Phase (d) & Telescope & Instrument & Wavelength coverage (\AA) & Reference    \\
 \hline
 \endhead
 \hline
 \endfoot
 \hline
 \endlastfoot
 \hline
 MJD  & Phase (d) & Telescope & Instrument & Wavelength coverage (\AA) & Reference    \\
 \hline
 55798.0 & $-$16.0 & Lijiang-2.4m & YFOSC & 3461 -- 8956 & \citet{spec_Lijiang-2.4m} \\
 55798.2 & $-$15.8 & Lick-3m  & KAST & 3416 -- 10278 & \citet{Nugent_1st_Lick_spec} \\
 55799.0 & $-$15.0 & Lijiang-2.4m & YFOSC & 3502 -- 8958 & \citet{spec_Lijiang-2.4m} \\
 55799.3 & $-$14.7 & UH88   & SNIFS & 3296 -- 9693 & \citet{spec_UH88} \\
 55800.2 & $-$13.8 & UH88   & SNIFS & 3296 -- 9693 & \citet{spec_UH88} \\
 55801.2 & $-$12.8 & UH88   & SNIFS & 3296 -- 9693 & \citet{spec_UH88} \\
 55802.3 & $-$11.7 & UH88   & SNIFS & 3296 -- 9693 & \citet{spec_UH88} \\
 55803.2 & $-$10.8 & UH88   & SNIFS & 3296 -- 9693 & \citet{spec_UH88} \\
 55804.2 & $-$9.8 & UH88   & SNIFS & 3296 -- 9693 & \citet{spec_UH88} \\
 55805.2 & $-$8.8 & UH88   & SNIFS & 3296 -- 9693 & \citet{spec_UH88} \\
 55806.2 & $-$7.8 & UH88   & SNIFS & 3296 -- 9693 & \citet{spec_UH88} \\
 55807.3 & $-$6.7 & UH88   & SNIFS & 3296 -- 9693 & \citet{spec_UH88} \\
 55808.2 & $-$5.8 & UH88   & SNIFS & 3296 -- 9693 & \citet{spec_UH88} \\
 55809.2 & $-$4.8 & UH88   & SNIFS & 3296 -- 9693 & \citet{spec_UH88} \\
 55811.4 & $-$2.6 & HST   & STIS & 1779 -- 24965 & \citet{spec_HST} \\
 55812.0 & $-$2.0 & Gemini-N  & GMOS & 3497 -- 9648 & \citet{spec_Gemini-N} \\
 55813.2 & $-$0.8 & UH88   & SNIFS & 3296 -- 9693 & \citet{spec_UH88} \\
 55814.2 & 0.2 & UH88   & SNIFS & 3296 -- 9693 & \citet{spec_UH88} \\
 55815.2 & 1.2 & UH88   & SNIFS & 3296 -- 9693 & \citet{spec_UH88} \\
 55816.2 & 2.2 & UH88   & SNIFS & 3296 -- 9693 & \citet{spec_UH88} \\
 55817.2 & 3.2 & UH88   & SNIFS & 3296 -- 9693 & \citet{spec_UH88} \\
 55817.7 & 3.7 & HST   & STIS & 1265 -- 24965 & \citet{spec_HST} \\
 55818.2 & 4.2 & UH88   & SNIFS & 3296 -- 9693 & \citet{spec_UH88} \\
 55821.2 & 7.2 & UH88   & SNIFS & 3296 -- 9693 & \citet{spec_UH88} \\
 55823.2 & 9.2 & UH88   & SNIFS & 3296 -- 9693 & \citet{spec_UH88} \\
 55826.2 & 12.2 & UH88   & SNIFS & 3296 -- 9693 & \citet{spec_UH88} \\
 55828.2 & 14.2 & UH88   & SNIFS & 3296 -- 9693 & \citet{spec_UH88} \\
 55829.0 & 15.0 & Gemini-N  & GMOS & 3497 -- 9643 & \citet{spec_Gemini-N} \\
 55830.2 & 16.2 & Keck1  & LRIS & 3227 -- 10242 & \citet{spec_Lick-3m} \\
 55831.2 & 17.2 & UH88   & SNIFS & 3296 -- 9693 & \citet{spec_UH88} \\
 55832.0 & 18.0 & Lijiang-2.4m & YFOSC & 3577 -- 8957 & \citet{spec_Lijiang-2.4m} \\
 55833.2 & 19.2 & UH88   & SNIFS & 3296 -- 9693 & \citet{spec_UH88} \\
 55835.3 & 21.3 & HST   & STIS & 1731 -- 10221 & \citet{spec_HST} \\
 55836.2 & 22.2 & UH88   & SNIFS & 3296 -- 9693 & \citet{spec_UH88} \\
 55838.2 & 24.2 & UH88   & SNIFS & 3296 -- 9693 & \citet{spec_UH88} \\
 55841.3 & 27.3 & HST   & STIS & 1738 -- 10221 & \citet{spec_HST} \\
 55855.2 & 41.2 & HST   & STIS & 1738 -- 10216 & \citet{spec_HST} \\
 55888.6 & 74.7 & UH88   & SNIFS & 3296 -- 9693 & \citet{spec_UH88} \\
 55891.7 & 77.7 & UH88   & SNIFS & 3296 -- 9693 & \citet{spec_UH88} \\
 55893.6 & 79.7 & UH88   & SNIFS & 3296 -- 9693 & \citet{spec_UH88} \\
 55896.6 & 82.6 & UH88   & SNIFS & 3296 -- 9693 & \citet{spec_UH88} \\
 55897.7 & 83.7 & Keck1  & LRIS & 3164 -- 10126 & \citet{spec_Lick-3m} \\
 55901.6 & 87.6 & UH88   & SNIFS & 3296 -- 9693 & \citet{spec_UH88} \\
 55903.6 & 89.6 & UH88   & SNIFS & 3296 -- 9693 & \citet{spec_UH88} \\
 55911.0 & 97.0 & XLT   & BFOSC & 3296 -- 9693 & \citet{spec_Lijiang-2.4m} \\
 55911.6 & 97.6 & UH88   & SNIFS & 3296 -- 9693 & \citet{spec_UH88} \\
 55913.5 & 99.5 & Lick-3m  & KAST & 3427 -- 10332 & \citet{spec_Lick-3m} \\
 55914.0 & 100.0 & WHT-4.2m  & ISIS & 3499 -- 9491 & \citet{WHT_spec_100d} \\
 55916.0 & 102.0 & WHT-4.2m  & ISIS & 3498 -- 9491 & \citet{PTF_1, PTF_2} \\
 55917.0 & 103.0 & WHT-4.2m  & ISIS & 3499 -- 9492 & \citet{PTF_1, PTF_2} \\
 55926.0 & 112.0 & Lijiang-2.4m & YFOSC & 3366 -- 9069 & \citet{spec_Lijiang-2.4m} \\
 55929.5 & 115.5 & Lick-3m  & KAST & 3426 -- 10170 & \citet{spec_Lick-3m} \\
 55944.5 & 130.5 & Lick-3m  & KAST & 3453 -- 10088 & \citet{spec_Lick-3m} \\
 55959.0 & 145.0 & Lick-3m  & KAST & 3497 -- 10000 & \citet{PTF_1, PTF_2} \\
 55980.4 & 166.4 & Lick-3m  & KAST & 3441 -- 10250 & \citet{spec_Lick-3m} \\
 55988.0 & 174.0 & WHT-4.2m  & ISIS & 3495 -- 9982 & \citet{spec_WHT-4.2m} \\
 56019.4 & 205.4 & Lick-3m  & KAST & 3438 -- 10324 & \citet{spec_WHT-4.2m} \\
 56040.4 & 226.4 & Lick-3m  & KAST & 3437 -- 10178 & \citet{spec_WHT-4.2m} \\
 56047.0 & 233.0 & Lijiang-2.4m & YFOSC & 3392 -- 9053 & \citet{spec_Lijiang-2.4m} \\
 56073.0 & 259.0 & WHT-4.2m  & ISIS & 3495 -- 9483 & \citet{spec_WHT-4.2m} \\
 56103.0 & 289.0 & WHT-4.2m  & ISIS & 3423 -- 10268 & \citet{spec_WHT-4.2m} \\
 56127.0 & 313.0 & P200   & DBSP & 3197 -- 10991 & \citet{PTF_1, PTF_2} \\
 56162.2 & 348.2 & Lick-3m  & KAST & 3487 -- 10240 & \citet{spec_WHT-4.2m} \\
 56194.2 & 380.2 & Keck1  & LRIS & 3232 -- 10268 & \citet{spec_Lick-3m} \\
 56277.0 & 463.0 & Lijiang-2.4m & YFOSC & 3379 -- 9337 & \citet{spec_Lijiang-2.4m} \\
 56778.5 & 964.5 & Keck1  & LRIS & 3074 -- 10320 & \citet{spec_Lick-3m+Keck1} \\
 56831.2 & 1017.2& LBT   & MODS1 & 3098 -- 10487 & \citet{spec_LBT} \\
\end{longtable}

\newpage

\begin{table}
 \centering
 \caption{Redshift values where 50 per cent of the simulated SNe were found to have CSM interaction. Strength is the strength of the \Halpha~line compared to the strength detected in SN 2015cp. Start shows how many days after the peak the interaction begins, and duration is in days as well. We fitted sigmoid functions to the results of each simulation in order to find the redshift where 50 per cent of the interactions were recovered, assuming the reference images were of the same depth as the ones used in ZTF or 0.5 or 1 mag deeper. These values are shown in $z_{50} \pm \sigma_{z_{50}}$, and $\chi^2_\text{red}$ shows the quality of the fit.}
 \begin{tabular}{cccclcrclcrclcr}
  \hline
  &&&& \multicolumn{3}{c}{ZTF references} && \multicolumn{3}{c}{0.5 mag deeper} && \multicolumn{3}{c}{1 mag deeper}\\
  strength & start & duration && $z_{50}$ & $\sigma_{z_{50}}$ & $\chi^2_\text{red}$ && $z_{50}$ & $\sigma_{z_{50}}$ & $\chi^2_\text{red}$ && $z_{50}$ & $\sigma_{z_{50}}$ & $\chi^2_\text{red}$\\
  \hline
  0.0 & -- & -- && 0.0038 & 0.0009 & 0.72 && 0.0042 & 0.0019 & 1.83 && 0.0046 & 0.0032 & 3.47\\
  0.1 & 100 & 100 && 0.0041 & 0.0008 & 0.83 && 0.0049 & 0.0011 & 1.82 && 0.0049 & 0.0033 & 3.41\\
  0.1 & 100 & 300 && 0.0050 & 0.0007 & 0.79 && 0.0055 & 0.0014 & 1.79 && 0.0059 & 0.0025 & 3.42\\
  0.1 & 100 & 500 && 0.0053 & 0.0006 & 0.77 && 0.0062 & 0.0011 & 1.76 && 0.0068 & 0.0018 & 3.44\\
  0.1 & 200 & 100 && 0.0045 & 0.0007 & 0.78 && 0.0050 & 0.0016 & 1.80 && 0.0054 & 0.0028 & 3.38\\
  0.1 & 200 & 300 && 0.0053 & 0.0006 & 0.78 && 0.0058 & 0.0014 & 1.79 && 0.0063 & 0.0022 & 3.44\\
  0.1 & 200 & 500 && 0.0056 & 0.0006 & 0.77 && 0.0066 & 0.0010 & 1.79 && 0.0073 & 0.0016 & 3.47\\
  0.1 & 300 & 100 && 0.0050 & 0.0006 & 0.80 && 0.0056 & 0.0012 & 1.80 && 0.0059 & 0.0023 & 3.43\\
  0.1 & 300 & 300 && 0.0052 & 0.0006 & 0.80 && 0.0061 & 0.0044 & 1.78 && 0.0067 & 0.0017 & 3.46\\
  0.1 & 300 & 500 && 0.0053 & 0.0005 & 0.82 && 0.0060 & 0.0010 & 2.03 && 0.0069 & 0.0016 & 3.53\\
  0.1 & 500 & 100 && 0.0045 & 0.0007 & 0.82 && 0.0048 & 0.0014 & 2.08 && 0.0054 & 0.0021 & 3.52\\
  0.1 & 500 & 300 && 0.0050 & 0.0005 & 0.83 && 0.0056 & 0.0010 & 2.09 && 0.0062 & 0.0017 & 3.53\\
  0.1 & 500 & 500 && 0.0050 & 0.0005 & 0.82 && 0.0057 & 0.0009 & 2.09 && 0.0063 & 0.0017 & 3.52\\
  1.0 & 100 & 100 && 0.0041 & 0.0007 & 0.86 && 0.0045 & 0.0013 & 2.10 && 0.0046 & 0.0027 & 3.58\\
  1.0 & 100 & 300 && 0.0091 & 0.0005 & 0.85 && 0.0096 & 0.0012 & 1.98 && 0.0108 & 0.0017 & 3.40\\
  1.0 & 100 & 500 && 0.0111 & 0.0004 & 0.84 && 0.0129 & 0.0008 & 2.08 && 0.0139 & 0.0013 & 3.50\\
  1.0 & 200 & 100 && 0.0067 & 0.0007 & 0.80 && 0.0077 & 0.0012 & 2.05 && 0.0083 & 0.0021 & 3.51\\
  1.0 & 200 & 300 && 0.0104 & 0.0005 & 0.93 && 0.0122 & 0.0008 & 2.06 && 0.0130 & 0.0014 & 3.31\\
  1.0 & 200 & 500 && 0.0116 & 0.0004 & 0.84 && 0.0129 & 0.0008 & 2.11 && 0.0140 & 0.0018 & 3.63\\
  1.0 & 300 & 100 && 0.0086 & 0.0004 & 0.78 && 0.0094 & 0.0008 & 1.95 && 0.0100 & 0.0013 & 3.62\\
  1.0 & 300 & 300 && 0.0105 & 0.0004 & 0.75 && 0.0118 & 0.0007 & 2.02 && 0.0128 & 0.0015 & 3.61\\
  1.0 & 300 & 500 && 0.0109 & 0.0004 & 0.78 && 0.0125 & 0.0008 & 2.11 && 0.0139 & 0.0013 & 3.76\\
  1.0 & 500 & 100 && 0.0081 & 0.0003 & 0.83 && 0.0090 & 0.0008 & 1.94 && 0.0094 & 0.0016 & 3.69\\
  1.0 & 500 & 300 && 0.0104 & 0.0003 & 0.78 && 0.0121 & 0.0007 & 2.10 && 0.0133 & 0.0012 & 3.79\\
  1.0 & 500 & 500 && 0.0105 & 0.0003 & 0.80 && 0.0122 & 0.0007 & 2.11 && 0.0135 & 0.0012 & 3.80\\
  10.0 & 100 & 100 && 0.0091 & 0.0014 & 0.93 && 0.0097 & 0.0028 & 2.13 && 0.0087 & 0.0058 & 3.45\\
  10.0 & 100 & 300 && 0.0259 & 0.0007 & 1.36 && 0.0299 & 0.0009 & 1.79 && 0.0328 & 0.0012 & 1.83\\
  10.0 & 100 & 500 && 0.0297 & 0.0005 & 1.21 && 0.0347 & 0.0007 & 1.83 && 0.0383 & 0.0010 & 1.84\\
  10.0 & 200 & 100 && 0.0204 & 0.0009 & 1.69 && 0.0223 & 0.0012 & 2.40 && 0.0217 & 0.0019 & 2.54\\
  10.0 & 200 & 300 && 0.0314 & 0.0004 & 1.45 && 0.0351 & 0.0007 & 1.98 && 0.0377 & 0.0010 & 2.40\\
  10.0 & 200 & 500 && 0.0326 & 0.0005 & 1.50 && 0.0374 & 0.0007 & 2.05 && 0.0415 & 0.0009 & 1.95\\
  10.0 & 300 & 100 && 0.0221 & 0.0023 & 1.21 && 0.0233 & 0.0011 & 2.01 && 0.0231 & 0.0019 & 2.71\\
  10.0 & 300 & 300 && 0.0311 & 0.0004 & 1.30 && 0.0350 & 0.0007 & 1.90 && 0.0375 & 0.0010 & 2.27\\
  10.0 & 300 & 500 && 0.0328 & 0.0004 & 1.42 && 0.0381 & 0.0007 & 2.22 && 0.0423 & 0.0009 & 2.21\\
  10.0 & 500 & 100 && 0.0199 & 0.0011 & 1.80 && 0.0224 & 0.0014 & 2.47 && 0.0226 & 0.0019 & 3.04\\
  10.0 & 500 & 300 && 0.0323 & 0.0004 & 1.23 && 0.0371 & 0.0007 & 2.12 && 0.0406 & 0.0009 & 2.19\\
  10.0 & 500 & 500 && 0.0323 & 0.0004 & 1.22 && 0.0373 & 0.0007 & 2.12 && 0.0407 & 0.0009 & 2.12\\
  \hline
 \end{tabular}
 \label{sim_z50_results}
\end{table}

\section{Inputs to \textsc{simsurvey} simulations}
\label{sec:simsurvey_inputs}

The specific inputs to \textsc{simsurvey} used in Section \ref{sim_obs} to determine the detection efficiencies for SN 2015cp-like interaction are listed here. 

\begin{itemize}
 \item Model: SN 2011fe + \Halpha~line (Section \ref{model_description}).
 \item Sky distribution: RA $\in$ [0$^{\circ}$, 360$^{\circ}$], Dec. $\geq -30^{\circ}$ (The area covered by ZTF, \citealt{ZTF_Surveys_Scheduler}).
 \item  Volumetric rate: The SN Ia rate is $2.4\times10^{-5}$ Mpc$^{-3}$ yr$^{-1}$ for $z \leq 0.1$ \citep{SNIa_rate}. \textsc{simsurvey} uses this to calculate the amount of SNe to generate at a given redshift interval.
 \item SN peak time distribution: $58\,195 \leq$ modified Julian date (MJD) $\leq 58\,487$ (between 18 March 2018 and 4 January 2019).
 \item Galactic extinction: dust maps from \citet{SFD98_dust_maps}.
 \item Host galaxy extinction: \citet{ccm89_extinction_law} extinction law, with \textit{E(B -- V)} drawn from an exponential distribution with exponent $\lambda=0.11$ \citep{EBV_simsurvey}, the same way as host extinction was added in the original \textsc{simsurvey} paper \citep{simsurvey_main}.
 \item Telescope specifications: ZTF P48 camera, 4$\times$4 grid of CCDs with four readout channels each, resulting in 64 separate output channels \citep{ZTF_Observing_System}.
 \item Survey plan: ZTF observation logs between $58\,197\leq$ MJD $\leq 59\,211$ (between 20 March 2018 and 28 December 2020), ensuring all simulated SNe are followed for a minimum of about 2 years after the peak.
\end{itemize}

\section{Late-time spectrum of SN 2020alm}
\label{spec_sec}
\begin{figure}
 \centering
 \includegraphics[width=\textwidth]{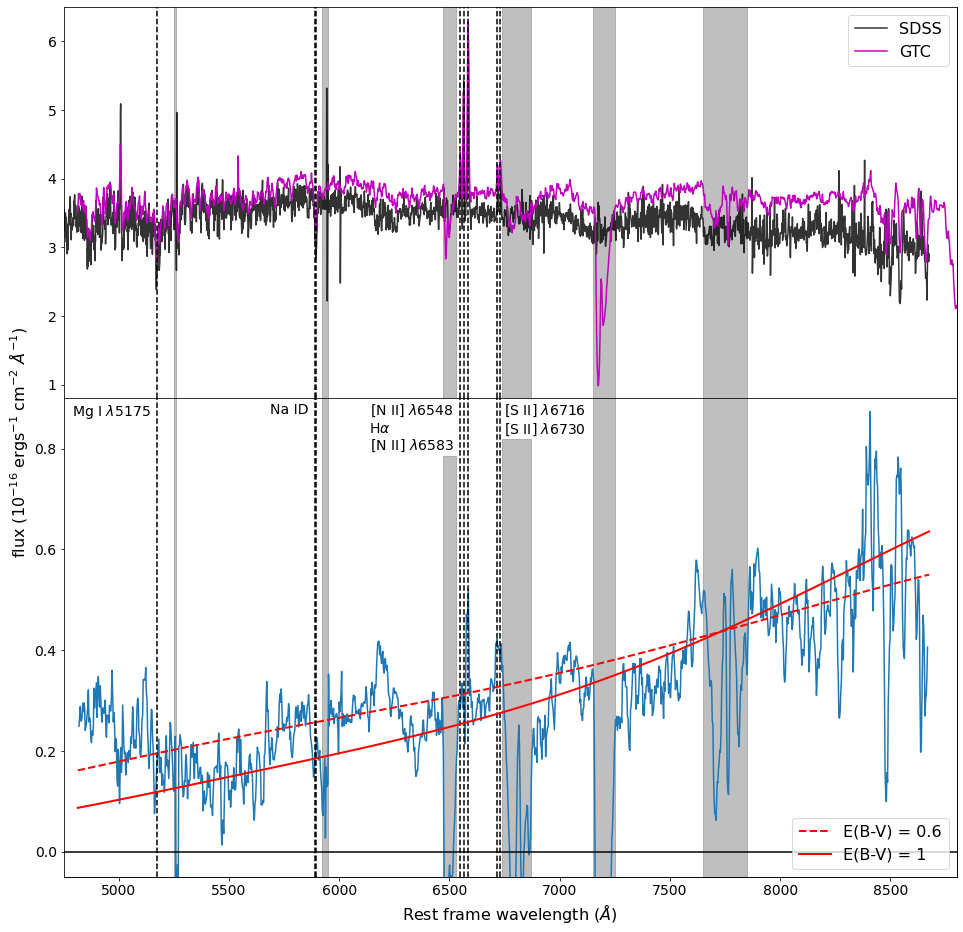}
 \caption{Spectrum of the late-time signal in SN 2020alm in its rest frame. The top panel shows the late-time spectrum obtained on 26 July 2023 using OSIRIS on the GTC, and the SDSS spectrum obtained in 2003. The bottom panel shows the late-time excess, obtained by subtracting the SDSS host galaxy spectrum from the observed late-time spectrum. A smoothed spectrum is shown in blue. The smoothing was done using a rolling kernel of size 5 to average over the values. The red lines are a simple TDE model with Milky Way and some amount of host galaxy extinction applied (the amount is shown in the legend), in order to get the approximate shape of the observed spectrum. Narrow emission and absorption lines that were notable in the unsubtracted spectrum are marked with dashed lines. The grey regions are affected by sky lines, and should be ignored.}
 \label{ZTF20aaifyfx_spec}
\end{figure}

After confirming the late-time detections in SN 2020alm to still be ongoing, we obtained a spectrum using OSIRIS on the GTC on 26 July 2023, 1277 days after the estimated peak date of the SN. As the observed spectrum was heavily dominated by the host galaxy, we subtracted a host galaxy spectrum taken by SDSS in 2003 to remove the host contamination. This was done after re-sampling the host spectrum to have the same wavelength spacing as the new spectrum. This left only the spectrum causing the late-time photometry detections, and is shown in Fig. \ref{ZTF20aaifyfx_spec}. The subtraction was confirmed to be successful by checking that the \MgI~${\lambda5175}$ and \NaID~absorption lines were reduced to the noise level, as these lines are not expected to be due to the late-time signal but purely from the host galaxy. Some of the host galaxy emission lines were not completely subtracted during this process, most noticeably [\NII]~${\lambda6583}$ and [\SII]~${\lambda\lambda6716,6730}$, but our resolution is inadequate to draw any conclusions from this.

The only explanation for the late-time detections that uses a second transient at the same location is a TDE. \citet{TDE_host_ext_range} shows that the intrinsic spectrum of a TDE is flat in this wavelength range, with sometimes some narrow emission lines. Therefore, we model a TDE spectrum as a line of constant flux density, and add Milky Way extinction \citep[using the SFD89 dust maps in the direction of the object; ][]{SFD98_dust_maps} and variable host extinction in an attempt to obtain the general shape of the observed spectrum. We find that $0.6\leq$ $E(B - V)$$_\text{host}\leq1$ mag is adequate to reproduce the general spectral shape and suggests that a TDE with approximately constant colour and moderate host extinction can explain the observed spectral excess for this event.

\section{Binned SMP light curves of the final candidates}
\begin{figure}
 \centering
 \includegraphics[width=\textwidth]{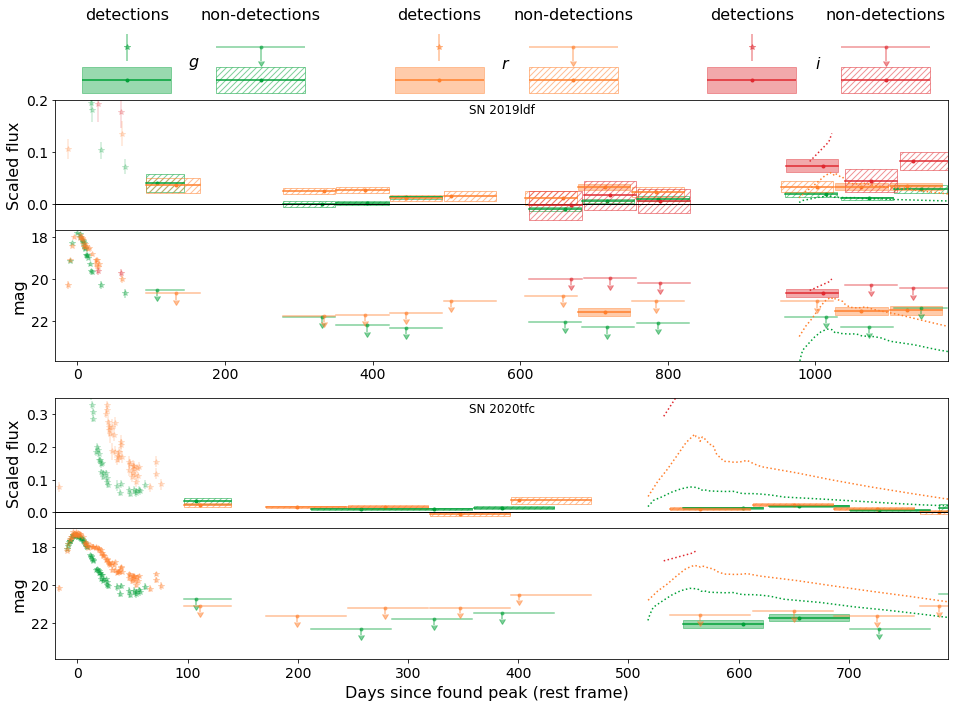}
 \caption{Similar to Fig.~\ref{candidates}, but showing the binned light curves generated using scene modelling photometry instead of forced photometry. As no bins $\geq5\sigma$ was recovered in SN 2018grt, it is not shown here. Both SN 2019ldf and SN 2020tfc do however still have robust detections in some bands. The best fitting alternate transients shown in dotted lines are the same as in Fig.~\ref{candidates}. The \textit{i}-band of SN 2020tfc is not shown as the background was not subtracted completely, resulting in a significant flux offset.}
 \label{final_candid_SMP}
\end{figure}
\end{appendix}
\end{document}